\documentclass[10pt,journal,compsoc]{IEEEtran}
\ifCLASSOPTIONcompsoc
  \usepackage[nocompress]{cite}
\else
  \usepackage{cite}
\fi
\usepackage{amssymb,amsmath,amsthm}
\usepackage{lastpage}
\usepackage{graphicx}
\usepackage{color}
\PassOptionsToPackage{hyphens}{url}\usepackage{hyperref}
\usepackage{xcolor} 
\hypersetup{
    colorlinks,
    linkcolor={black},
    urlcolor={black},
    citecolor={black},
} 
\usepackage[font={footnotesize,bf}]{subcaption}
\usepackage[font={footnotesize,bf}]{caption}
\usepackage{inputenc}
\usepackage{algorithm}
\usepackage{algpseudocode}
\usepackage{multirow}
\usepackage{comment}
\usepackage{booktabs}
\usepackage{url}
\usepackage{cleveref}
\usepackage[]{todonotes}
\presetkeys%
    {todonotes}%
    {inline}{}
\usepackage{setspace}
\usepackage{relsize}
\usepackage{makecell}
\usepackage{tabularx} 
\usepackage{enumitem}
\usepackage{adjustbox,lipsum}
\usepackage[normalem]{ulem}

\DeclareMathOperator{\sign}{sign}

\newtheorem{lemma}{Lemma}
\newtheorem{corollary}{Corollary}
\newtheorem{definition}{Definition}

\begin{document}
\title{Max-Min Fair Resource Allocation in Millimetre-Wave Backhauls}
%
%
%
\author{Rui Li and Paul Patras
\thanks{The authors are with the School of Informatics, University of Edinburgh, UK.}%
\thanks{A preliminary version of this paper appeared in ACM HotWireless 2016 \cite{Li:2016:HotWireless}.}
}

%
\IEEEtitleabstractindextext{
\begin{abstract}
5G mobile networks are expected to provide pervasive high speed wireless connectivity and support increasingly resource intensive user applications. Network hyper-densification therefore becomes necessary, though connecting to the Internet tens of thousands of base stations is non-trivial, especially in urban scenarios where optical fibre is difficult and costly to deploy. 
The millimetre wave (mm-wave) spectrum is a promising candidate for inexpensive multi-Gbps wireless backhauling, but exploiting this band for effective multi-hop data communications is challenging. In particular, resource allocation and scheduling of very narrow transmission/ reception beams require to overcome terminal deafness and link blockage problems, while managing fairness issues that arise when flows encounter dissimilar competition and traverse different numbers of links with heterogeneous quality.
In this paper, we propose {\sc WiHaul}, an airtime allocation and scheduling mechanism that overcomes these challenges specific to multi-hop mm-wave networks, guarantees max-min fairness among traffic flows, and ensures the overall available backhaul resources are fully utilised. We evaluate the proposed {\sc WiHaul} scheme over a broad range of practical network conditions, and demonstrate up to 5$\times$ individual throughput gains and a five-fold improvement in terms of measurable fairness, over recent mm-wave scheduling solutions.
\end{abstract}
}
\maketitle

\IEEEraisesectionheading{\section{Introduction}}

Market surveys confirm that the number of mobile subscriptions and the popularity of bandwidth-intensive applications (including ultra high-definition video and virtual/ augmented reality) continue to grow at an unprecedented pace \cite{Ericsson:2017}. In response to the accelerating traffic demands, carriers are offering flat-rate unlimited data plans~\cite{forbes:2017}, which requires to substantially extend the capabilities of current mobile network infrastructure. 
Cell densification is a first step \cite{Bhushan:2014}, but this entails revisiting existing backhauling practices, to be able to transfer vast volumes of data between the access and core networks.
In particular, the cost of deploying traditional, fibre-based backhauls surges with network density, whilst the reconfiguration of such solutions is limited. 
Wireless alternatives have been thus far confined to the microwave spectrum (0.3--30GHz), which is of restricted capacity and already overcrowded with numerous applications, including \mbox{Wi-Fi}, Bluetooth, digital video broadcast (DVB), cellular access, radar, and machine-to-machine (M2M) communications.
 
The millimetre-wave (mm-wave) band (30--300GHz) is in contrast largely underutilised and exposes considerably wider spectral resources that could support an order of magnitude higher data rates\cite{rappaport:2013}. As a result, regulatory bodies such as Ofcom in the UK are encouraging nation-wide 5G trials in the mm-wave band~\cite{ofcomm:2018}, and industry stakeholders have begun collaborating on building multi-Gbps millimetre-wave backhaul solutions in urban areas (see e.g. the involvement of Qualcomm and Facebook in the Terragraph project\cite{facebook:terragraph,qualcomm:2018}).  
3GPP further promotes mm-wave technology through the specification of 5G new radio (NR) in release 15 of the mobile broadband standard~\cite{3GPP-NR}, with the first systems already being prototyped~\cite{qualcomm:2017:cellran}.
Harnessing the potential of mm-wave bands is however only possible with highly-directional beamforming using multiple antennas and phase arrays, which mitigates the severe signal attenuation characteristic to these frequencies. Previous research efforts provide sufficient evidence of the effectiveness of this approach and the feasibility of mm-wave communications for small cell backhauling~\cite{hur2013millimeter,alexandropoulos2017position}.
Directionality intrinsically eliminates interference and enables better spatial reuse, though introduces the risk of \emph{link blockage}, due to moving obstacles, and \emph{terminal deafness}, i.e. receivers can hardly be aware of transmitters, unless their beams are mutually aligned~\cite{roh:2014}.
The latter is particularly problematic in deployments with small form factor base stations (such as in urban lamppost based infrastructure) that serve large numbers of end-users over \mbox{Wi-Fi}/cellular and communicate with gateways using single mm-wave transceivers, over multiple hops. 

\begin{figure}[!t]
	\begin{center}
 	\includegraphics[width=0.9\columnwidth]{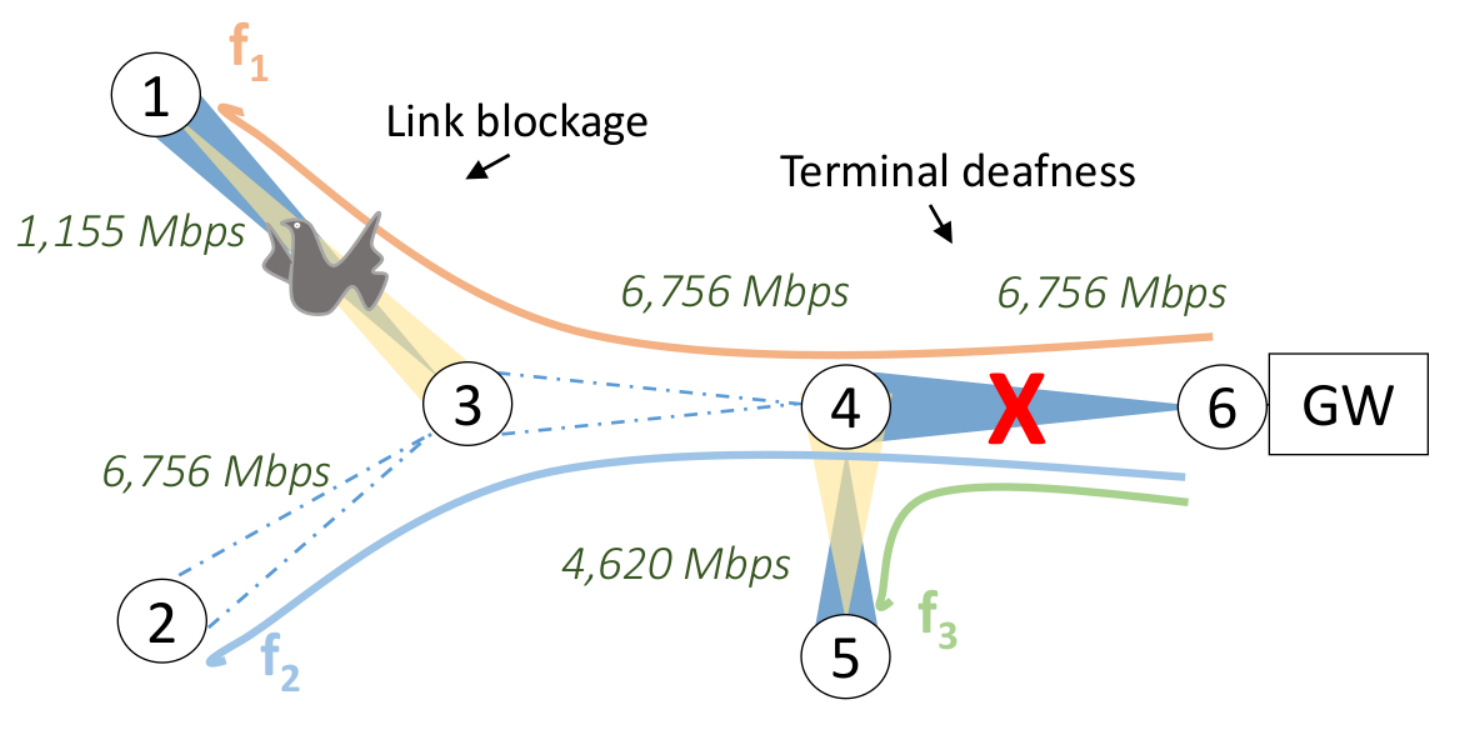}
 	\caption{Mm-wave backhaul with 6 base stations. Three aggregate flows traverse the network in the downlink direction (paths shown with continuous lines). TX/RX beams shown with  dark/light shades, possible beams with dashed lines. Link 6-4 subject to terminal deafness, 3-1 partially blocked. Link bit rates labelled.}
 	\label{fig:simpletopo}
	\end{center}
\vspace*{-0.5em}
\end{figure}

In this new setting, the key challenge is deciding at each base station to which neighbour to transmit or receive from, when, and for how long, so as to fully utilise the available resources. This is effectively a medium access scheduling task constrained by the demand of the flows traversing the network, fairness requirements, and physical link properties at any given time. To better appreciate the difficulty of this task, consider the example scenario illustrated
in Fig.~\ref{fig:simpletopo}, where 6 base stations communicate over mm-wave links with a wired gateway. Here, three high volume traffic flows are relayed by intermediary hops from the gateway towards base stations 1, 2, and 5 respectively. 
Station 6 is locked out when attempting to transmit to station 4, if this station has its TX/RX beams steered towards station 5. In addition, the communication between stations 1 and 3 is partially blocked by a moving object, resulting in link quality degradation. 
Further, the three backlogged traffic aggregates traversing the backhaul in this example are relayed over different number of hops, and encounter different level of competition on heterogeneous links. Therefore, the airtime allocation strategy will impact on the distribution of resources and lead to fairness issues and/or sub-optimal network utilisation, unless all these aspects are carefully considered. 

Indeed, max-throughput strategies favour large volume flows traversing high capacity links, while round-robin schemes that allocate equal airtime are proportionally fair, but lead to wastage of network resources, as summarised in Table~\ref{tab:examplerates}. In the table we also indicate the performance of the {\sc WiHaul} max-min fair backhauling scheme, which we propose in this paper. This clearly yields the smallest level of unfairness, as quantified with the Gini coefficient~\cite{gini:1921}, and only 10\% lower total network throughput as compared to the greedy max-throughput strategy, which allocates all resources to a single flow.

\begin{table}[t!]
\begin{center}
\adjustbox{width=\columnwidth}{
 \begin{tabular}{c | c c c |c |c} 
Scheme &Flow 1 & Flow 2 & Flow 3 & \begin{tabular}{@{}c@{}} Total \\ throughput\end{tabular} & \begin{tabular}{@{}c@{}} Gini\\  coefficient\end{tabular}\\
 \hline\hline
Max-throughput &0Mbps	&3,378Mbps	&0Mbps	&3,378Mbps &0.6667 \\
\hline
 \begin{tabular}{@{}c@{}}Round-Robin \\ (equal airtime)\end{tabular} &289Mbps	&1,126Mbps		&770Mbps	&2,185Mbps	&0.2554\\
  \hline
 \begin{tabular}{@{}c@{}}\textbf{Proposed WiHaul} \\ \textbf{(max-min)}\end{tabular} & \textbf{763Mbps}	&\textbf{763Mbps}	&\textbf{1,504Mbp}s	&\textbf{3,030Mbps} &\textbf{0.1630}\\
 \bottomrule
\end{tabular}
}
 \caption{Rate distribution, total throughput, and unfairness measure with different resource allocation schemes for the topology shown in Fig.~\ref{fig:simpletopo}. Numerical example.}
 \label{tab:examplerates}
 \end{center}
 \end{table}

Backhaul solutions designed with legacy multi-hop wireless technology operating in sub-6GHz bands are inappropriate, given the unique properties of mm-wave communications. As the infrastructure has commercial value, it is essential to ensure resources are not left underutilised, while customers remain satisfied with the level of service provided. Several 5G standards define carrier-grade mechanisms that allow for precise scheduling (e.g. 3GPP NR~\cite{3GPP-NR} and IEEE 802.11ad\footnote{We note that, although the IEEE 802.11ad is primarily intended for single-hop wireless local area networks, this protocol could also be used for multi-hop solutions in unlicensed bands, e.g. 60GHz, serving community networks.} with Service Period operation~\cite{802.11ad}), yet the \emph{airtime allocation and scheduling tasks, which are crucial for backhauling, are left open to implementation}.

In this paper we jointly solve the airtime allocation and per-link scheduling of \emph{aggregate} traffic flows, i.e. flow bundles that originate/terminate at the same base station,\footnote{Hereafter, whenever there is no scope for confusion, we use the terms `flow' and `aggregate flow' interchangeably.} which traverse multi-hop mm-wave backhauls. We focus on allocating
resources at the medium access control (MAC) layer for general mm-wave systems. We do not make contributions in terms of PHY layer optimisation and argue that aspects including power allocation, codebook design, or beamform training can be largely decoupled from MAC operation; however, we explicitly take into account the distinct features of mm-wave technology, i.e. terminal deafness and susceptibility to link blockage, as well as realistic heterogeneous traffic demand regimes. Our goal is to \emph{achieve a good balance between overall network throughput performance and inter-flow fairness}. That is, the revenue obtained from operating backhauls can be maximised, whilst aggregate flows encountering low capacity links and/or increased competition are not unnecessarily throttled (high quality of service). Our focus is on providing small cell backhauling that can cater for real-time applications where latencies below tens of milliseconds are not perceivable by the user, instead of minimising latency, as required by ultra low-latency applications.
As such, we make the following \textbf{key contributions}:

\begin{enumerate}
\setlength{\itemsep}{0pt}
 \item We cast backhaul resource allocation as a max-min\footnote{
 We work with the max-min criterion instead of the popular Jain's fairness index, as we aim to avoid resource under-utilisation incurred when equalising throughputs. Instead, we seek to fulfil flow demands in increasing order, while sharing remaining network  capacity among  flows  with  higher  demands. In the absence of an established quantitative measure of max-min fairness, we work with the notion of economic inequality (i.e. the  Gini coefficient~\cite{gini:1921}) and extend a generic fairness model~\cite{lan:2010} to further quantify max-min fairness.} optimisation problem with mm-wave specific terminal deafness and potential secondary interference, and traffic demand constraints. We demonstrate that a max-min fair solution exists and it is unique in scheduled-based multi-hop mm-wave networks.
 
 \item We propose {\sc WiHaul}, a backhauling scheme comprising \emph{(i)} a progressive filling algorithm that solves the max-min optimisation problem and computes per-hop airtime shares for each aggregate flow, and \emph{(ii)}~a light-weight scheduling protocol that works on top of any time-division  multiplexing (TDM) protocol for mm-wave systems, enforces the computed airtimes, and coordinates multi-hop transmissions, enabling spatial~reuse.
 
 \item We implement {\sc WiHaul} in the NS-3 simulator, building on preliminary mm-wave PHY measurements and incorporating the IEEE 802.11ad specification, with extended functionality for multi-hop settings. Although this does not bear features specific to cellular systems, the MAC operation in the time domain is largely similar, thus the results obtained are relevant to such systems as well. We evaluate the performance of our solution over different network topologies, link dynamics, routing paradigms, and traffic regimes. 
 We demonstrate up to 5-fold throughput and fairness gains over previously proposed mm-wave access schemes.

\end{enumerate}

\section{Related Work}
\label{sec:relatedworks}

To the best of our knowledge, the proposed {\sc WiHaul} scheme is the first to perform airtime allocation in mm-wave backhauls, explicitly addressing the distribution of flow rates and quantifying max-min fairness. In what follows, we review related work of direct relevance to our contribution.

\textbf{Mm-wave Characterisation \& Exploitation:} 
Recent empirical studies confirm the millimetre-wave band (30--300GHz) will be able to support multi-Gbps link rates \cite{rappaport:2013}. Hence, it becomes a promising candidate to accommodate bandwidth intensive small-cell wireless backhauling solutions~\cite{MacCartney:2014}. Channel measurement efforts also confirm that beamforming necessary to mitigate attenuation in mm-wave bands drastically reduces interference, and links can often be regarded as pseudo-wired~\cite{Singh2011Interference}. Wang \emph{et al.} propose a code-book based beamforming protocol to setup multi-Gbps mm-wave communication links~\cite{wang2009bfcodebook}. Hur \emph{et al.} design a beam alignment mechanism for mm-wave backhauling scenarios, tackling the effects of wind-induced beam misalignment~\cite{hur2013millimeter}. With mandatory use of beamforming, however, terminal deafness becomes a key challenge when scheduling transmissions/receptions~\cite{Nitsche:2014}. The throughput and energy consumption characteristics of different mm-wave bands are studied in \cite{Mesodiakaki:2016}. While we do not explicitly address energy efficiency aspects in our work, we recognise that a certain degree of energy efficiency can be inherently achieved through optimal airtime allocation and scheduling, which is at the core of our work.

\textbf{Medium Access \& Scheduling in Mm-wave Networks:} 
Medium Access Control protocols for mm-wave communications can be grouped into two main classes -- contention-based and (pseudo-)scheduled.
The IEEE 802.11ad standard~\cite{802.11ad} specifies both contention-based and Service Period (SP) driven (scheduled) channel access mechanisms for communications in the unlicensed 60GHz band. 
Building upon 802.11ad, the 802.11ay draft aims to achieve link rates of up to 100Gbps, by employing a number of enhancements, including 4-stream MIMO~\cite{80211ayStatus}. On the other hand the 3GPP New Radio (NR) specification extends the LTE numerology by allowing different types of sub-carrier spacing and slot lengths~\cite{3GPP-NR}. The 10ms frame structure of LTE with 1ms subframes is preserved. It is worth noting that both IEEE and 3GPP standards leave open the airtime allocation and multi-hop transmission coordination tasks.

Hemanth and Venkatesh analyse the performance of the 802.11ad SP mechanism in terms of frame delay~\cite{hemanth:2016}. 
Several works build upon the 802.11ad standard and specify MAC protocol improvements for single-hop WLANs \cite{chandra:2014,sim:2016,chen:2013}. Chandra \emph{et al.} employ adaptive beamwidth to achieve improved channel utilisation~\cite{chandra:2014}. Sim \emph{et al.} exploit dual-band channel access to address terminal deafness and improve throughput~\cite{sim:2016}. Optimal client association and airtime allocation is pursued in \cite{facchi:2017} to maximise the utility of enterprise mm-wave deployments.

A directional cooperative MAC protocol is introduced in~\cite{chen:2013}, where user devices select intermediate nodes to relay the packets to the AP, in order to establish multi-hop paths that exhibit higher signal-to-noise ratio (SNR) than direct links. Mandke and Nettles propose a dual-band architecture for multi-hop 60GHz networks where scheduling and routing decisions are communicated at 5.2GHz~\cite{mandke:2010}. Based on their feasibility study of in-band wireless backhauling, Taori \emph{et al.} present a qualitative scheduling framework for inter-base station communications~\cite{taori2015point}. This resembles closely the Type 2 TDD scheme of LTE, with the difference that the authors apply it to in-band backhauling scenarios, whereas in the LTE standard this is specified for cellular access only. Despite considering the implications of terminal deafness, these designs do not tackle the airtime allocation problem. 
Relay selection so as to overcome blockage and scheduling in mm-wave backhauls is tackled in \cite{hu2017}, with the aim of maximising throughput. However, neither airtime allocation nor fairness are taken into account.

Distributed opportunistic transmission schemes for multi-hop scenarios have been proposed to achieve network-wide scheduling~\cite{sim2016learning,singh2010distributed}. 
MDMAC operates with a slotted channel whereby a station's transmission can occupy one or multiple slots, but the slot duration remains fixed for all participants (20$\mu$s by default), which may harm efficiency~\cite{singh2010distributed}. 
Unslotted approaches named (Bin)DLMAC are introduced in~\cite{sim2016learning} to improve protocol efficiency and 'learn' when to transmit in the presence of terminal deafness.
Both schemes do not explicitly consider inter-flow fairness, as each node seeks to transmit as much as possible. Our results confirm that this leads to poor performance for flows encountering lower capacity links. 

Su and Zhang solve optimal network throughput allocation heuristically in multi-channel settings, without fairness guarantees~\cite{su:2009}. Ford \emph{et al.} target sum utility maximisation in self-backhauled mm-wave setting~\cite{ford:2015}. Seminari \emph{et al.} formulate the sharing of mm-wave backhauls as a one-to-many matching game, seeking to maximise the average sum rate~\cite{seminari:2017}.
Zhu \emph{et al.} propose a maximum independent set (MIS) based scheduling algorithm to maximise QoS in mm-wave backhauls~\cite{zhu2016qos}. Similarly, Niu \emph{et al.} propose MIS based scheduling that aims to minimise the energy consumption~\cite{niu2017energy}. A joint scheduling and power allocation problem is also solved with MIS in \cite{li2017joint}. In this body of work scheduling is performed with the explicit goal of achieving concurrent transmissions among non-interfering links. The {\sc WiHaul} mechanism we propose allows for concurrent transmissions by default. Moreover, {\sc WiHaul} not only improves throughput performance, but also explicitly addresses fairness, while we take into account all flow demands, link rates, and the level of competition among them. In particular, we address airtime allocation and scheduling in multi-hop mm-wave networks using the max-min fairness criterion.

\textbf{Max-Min Fairness in Multi-hop Wireless Networks:} Bertsekas and Gallager consider max-min fairness for flow control in wired networks \cite{Bertsekas:1992} and  subsequently Le Boudec and Radunovic demonstrate this is a geometric property of the set of feasible allocations~\cite{Radunovic:2007}. The 802.11 rate region is proven log-convex, and station attempt probabilities and burst sizes in 802.11 mesh networks are derived for max-min fair regimes in~\cite{Leith:2012}. This however only holds in multi-channel mesh topologies where stations employ multiple interfaces, which is impractical with small form factor mm-wave devices equipped with a single interface. 
Wang \emph{et al.} argue that channel time rather than flow rate should be used with the max-min allocation criterion in wireless multi-hop networks and accordingly propose a new definition of max-min fairness~\cite{Wang:2008}. Unfortunately, under this definition, flows traversing more hops will, by design, obtain considerably smaller throughput than those close to gateways. This implies inferior service performance for distant users, hence the approach is ill-suited to the carrier-grade backhauls.

Lan \emph{et al.} propose a unified fairness measure that enables to explicitly quantify max-min fairness, which is largely perceived as qualitative \cite{lan:2010}.
We use their general measure of fairness to derive a max-min fair metric and evaluate the gains achieved by our proposal. To add further perspective, in our evaluation we also resort to 
economic notions of inequality, i.e. the 
Gini coefficient \cite{gini:1921}.

\section{System Model}
\label{sec:sysmodel}

We focus on dense mobile broadband deployments where $B$ fixed base stations provide wireless access to mobile users with different traffic demands. While serving a number of smart devices that consumes or generate data flows, base stations are connected via mm-wave links to wired Internet gateways, possibly over multiple hops. 

\textbf{PHY Layer Considerations:}
Although PHY layer optimisation is outside the scope of our work, we briefly summarise the channel model considered. We assume each backhaul node employs $N$ TX/RX antennas and adopt the mm-wave MIMO channel model proposed in \cite{alkhateeb2014channel},
where hybrid analogue/digital pre-coding is employed. By \cite{alkhateeb2014channel}, the channel is subject to limited scattering and geometric models are generally applicable~\cite{rappaport2013broadband, rappaport2012cellular}. The channel matrix can be expressed as
\begin{align}
    \textbf{H} = \frac{N}{\sqrt{\overline{P_L}}} \sum_{l=1}^{L}\alpha_l \textbf{a}_{rx}(\theta_l^{AOA}) \textbf{a}_{tx}^{H}(\theta_l^{AOD}),
\end{align}
where $\overline{P_L}$ denotes the average path-loss between a transmitter and receiver, $L$ is the number of scatterers, and $\alpha_l$ is the complex gain of the $l$-th channel, following the Rayleigh distribution $\alpha_l \sim {\sc N} (0, \overline{P_R}), \forall l \in \{1,2,...,L\}$. $\overline{P_R}$ is the average power gain. $\theta_l^{AOD} \in [0, 2\pi]$ and $\theta_l^{AOA} \in [0, 2\pi]$ denote the azimuth angles of departure and arrival, respectively, and $\textbf{a}_{tx}(\theta_l^{AOD})$ and $\textbf{a}_{rx}(\theta_l^{AOA})$ are the antenna array response vectors at the transmitter and the receiver. While extensions to 3D beamforming is possible~\cite{ayach:2014}, we focus here on horizontal 2D beamforming and neglect the elevation angle. Assuming uniform linear arrays, the antenna response vector can be written as:
\begin{align}
    \textbf{a}_{tx}(\theta_l^{AOD}) = \frac{1}{\sqrt{N}} [1, e^{j(2\pi/\lambda) d \sin (\theta_l^{AOD})}
    ,...,\notag\\
    e^{j(N - 1)(2\pi/\lambda) d \sin (\theta_l^{AOD})} ]^T,
\end{align}
where $\lambda$ is the wavelength, and $d$ is the distance between antenna elements. The response vector of the receiver antenna array, i.e. $\textbf{a}_{rx}(\theta_l^{AOA})$ has a similar form.

According to \cite{alkhateeb2014channel}, with efficient design of the pre-coders ($\textbf{F}_{BB}$ for baseband and $\textbf{F}_{RF}$ for radio frequency -- RF) and combiners ($\textbf{W}_{BB}$ for baseband and $\textbf{W}_{RF}$ for RF), the achievable rate of the MIMO system is formulated as:

\begin{align}
\begin{split}
    R = \log _2 |\textbf{I}_{N} + \frac{P}{N_S} \textbf{R}_n^{-1}\textbf{W}_{BB}^H \textbf{W}_{RF}^H \textbf{H}\textbf{F}_{RF} \textbf{F}_{BB}\notag\\ \textbf{F}_{BB}^H \textbf{F}_{RF}^H \textbf{H}^H \textbf{W}_{RF} \textbf{W}_{BB}|,
\end{split}
\end{align}

\noindent where the post-processing noise co-variance matrix $\textbf{R}_n$ is given by $\textbf{W}_{BB}^H \textbf{W}_{RF}^H \textbf{W}_{RF} \textbf{W}_{BB}$.

\begin{figure}[!t]
	\begin{center}
 	\includegraphics[width=0.8\columnwidth]{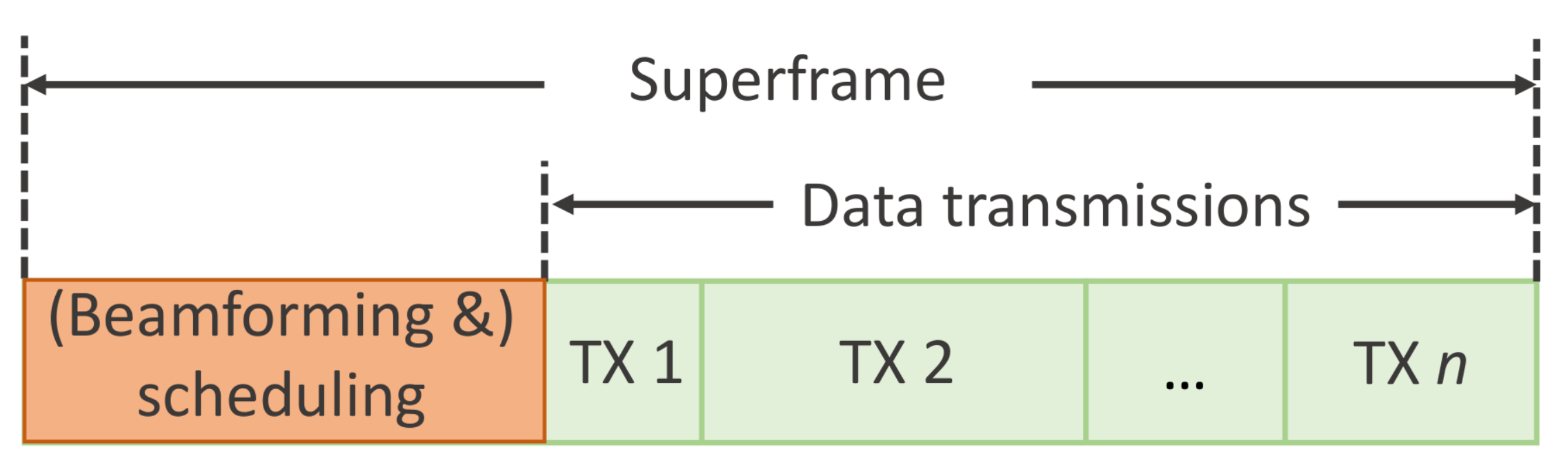} 
 	\caption{TDM superframe structure observed bv {\sc WiHaul}. Beamform training, scheduling and control message exchange take place periodically at the start. Data transmissions (possibly of different durations) follow. The structure can equally apply to 3GPP NR and IEEE 802.11ad with SPs.}
 	\label{fig:framestruct}
	\end{center}
\end{figure}

\textbf{MAC Paradigm:} We target mm-wave systems where channel multiplexing is performed following time division principles. As such, our solution is applicable to both TDMA-based cellular backhaul scenarios and single-/multi-hop deployments based on the IEEE 802.11ad standard~\cite{802.11ad} working with SPs, e.g. in rural and community networks. With these in mind, we address rigorously the airtime allocation and TX/RX beam scheduling in multi-hop backhaul networks.
{\sc WiHaul} observes a periodic superframe/beacon interval structure where beamform training information is exchanged and TX/RX scheduling is performed at the start of a superframe, following which link transmissions take place as per computed schedules, as depicted in Fig.~\ref{fig:framestruct}.

\begin{figure}[!b]
\vspace*{-1em}
	\begin{center}
 	\includegraphics[width=\columnwidth]{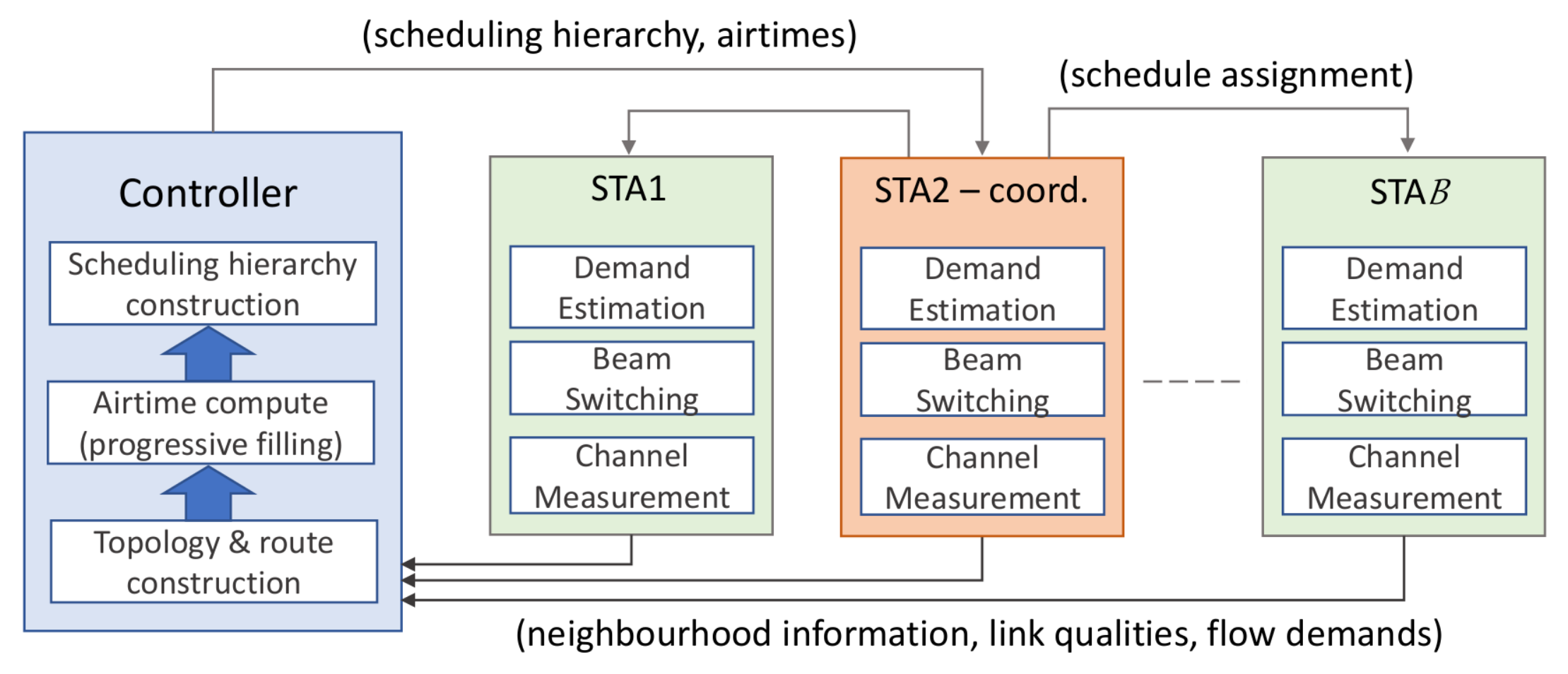} 
 	\caption{High-level overview of the envisioned system. {\sc WiHaul} runs on the controller and computes flow airtime allocations and schedules, based on topology information and paths computed by routing logic. Scheduling hierarchy and airtimes sent to a scheduling coordinator, which dictates the TX/RX timing to backhaul nodes.}
 	\label{fig:system}
	\end{center}
\end{figure}

\textbf{Centralised control:} We envision a centralised architecture, whereby a controller has full knowledge of the network topology, periodically collects link rate and flow demand information, and subsequently performs airtime allocation and beam scheduling through the solution we introduce in this work. In practice, centralised control is achievable through software defined network (SDN) primitives \cite{haleplidis:2015}, for instance running OpenFlow~\cite{haque2016sdnsurvey} over a dedicated narrow-band low frequency channel. Similar approaches out-of-band control schemes have been previously used in wide-spectrum \cite{Giannoulis:2013:INFOCOM} and multi-hop 60GHz networks~\cite{mandke:2010}.  We assume the controller is also responsible for computing paths $p_k$ for all flows $k$ traversing the backhaul, which is orthogonal to the problem we attack and thus not explicitly considered herein. This is aligned with previous work on mm-wave backhauling where path computation and link scheduling are dealt with separately~\cite{seppanen:2016}. We give an overview of the overall envisioned system in Fig.~\ref{fig:system}.

Our objective is to allocate the airtime resources available on the mm-wave backhaul links to aggregate traffic flows and co-ordinate transmissions among base stations. Flows either enter the network via gateways, are relayed by intermediary hops, before reaching the end users (downlink), or originate at different base stations and are forwarded externally by the gateways (uplink). The problem we pursue is challenging and fundamentally different to previous efforts in multi-hop wireless networks (e.g. \cite{Wang:2008}), since the backhaul system is prone to terminal deafness and a receiver may experience secondary interference when situated in the range and on the direction of another active beam. 
Since we consider deployments with small form factor base stations equipped with a single mm-wave interface, intra-flow competition occurs and fairness issues arise as flows are relayed by base stations, unlike in multi-radio mesh networks~\cite{Leith:2012}. Meanwhile, concurrent transmissions on non-interfering links is feasible, which allows for spatial reuse and appropriate network utilisation at a lower cost.


\section{Problem Formulation}
\label{sec:problemformulation}  

The problem we seek to solve is how to distribute mm-wave backhaul resources among flows, such that network utilisation is maximised while flows with lower demand or originating/terminating further away from the gateway are not throttled. Our focus is on the MAC layer and we assume
PHY layer aspects such as power allocation, code-book design, or beamform training can be dealt with separately. This assumption is reasonable, because PHY optimisation will ultimately result in different capacity constraints imposed on the MAC layer. Mm-wave specific aspects such as link blockage are inherently captured in our formulation through constraints. We further take into account potential link-blockage conditions when allocating resources, circumventing these as we explain below.
We work with the \emph{max-min fair} criterion \cite{Bertsekas:1992}, seeking to ensure flow demands are fulfilled in increasing order where possible, whilst any remaining network capacity is shared among flows with higher demands. That is, we aim to maximise the end-to-end throughput $r_k$ of each aggregate flow $k$, subject to individual demands $d_k$, whilst any flow allocation increase would not harm others with already smaller or equal throughputs. We denote $c_{i,j}$ the maximum achievable data rate between an $(i,j)$ base station pair and work with aggregate data traffic flows between base stations and the gateway. To the best of our knowledge, max-min fair resource allocation in mm-wave backhauls, which present unique terminal deafness challenges, has not been considered previously.

Formally, by denoting $\cal{F}$ the set of flows traversing the backhaul, $p_k$ the path of flow $k$, i.e. the sequence of links this follows from source to destination (within the backhaul), and considering flow $k$ is assigned airtime $t_{k,i,j}$ on link $l_{i,j}$, we want to find the vector
\[
 \mathbf{t} := \{t_{k,i,j} | k \in {\cal{F}}, l_{i,j} \subset p_k\} 
\]
that achieves max-min fair allocation of flow throughputs. This requires to solve the following optimisation problem in an iterative fashion:

\begin{align}
 &\mathbf{t}^* = \mathop{\bigcup}\limits_{k \in \cal{F_A},\ \cal{F_A}\subset \cal{F} }  \mathop{\arg\max}\limits_{\mathbf{t}}\ \min_{k} r_k  \label{eq:objective}\\
 \text{s.t.}\ & r_k \leq d_k, \forall k \in {\cal{F_A}}, \label{eq:demand_constraint} \\
 & {\bgroup
  \sum_{s_{k,i,j} \in C_q}  \frac{r_k}{c_{i,j}} \leq 1 - \tau, \forall k \in {\cal{F_A}}, \forall C_q \in \cal{C}.\label{eq:clique_constraint}\egroup}
\end{align}

\noindent In the above, $\cal{F_A} \subset \cal{F}$ denotes the set of flows that have not yet been allocated end-to-end resources (active flows) and~(\ref{eq:demand_constraint}) represents a \emph{demand constraint} that ensures any allocated flow rate does not exceed the corresponding demand, so that no resources will be left unused.
$s_{k,i,j}$ in~(\ref{eq:clique_constraint}) represents the segment of flow~$k$ traversing link $l_{i,j}$ for which we seek to allocate $t_{k,i,j}$ airtime. 

As single transceiver stations can only send to, or receive from one neighbour at a time, we construct a conflict graph $G(V,E)$, where a flow segment corresponds to a vertex $v \in V$. An edge $e \in E$ exists between any two vertices, if the corresponding flow segments cannot be simultaneously active, either because they traverse the same node or because they may cause secondary interference onto one another, due to beam alignment and transmission range. $C_q$ denotes a \emph{clique}, which follows the definition we give below.

\begin{definition}
 A `clique' is the set of all flow segments that cannot be active simultaneously.
\end{definition}

\begin{figure}[!t]
	\begin{center}
			 \includegraphics[width=0.7\columnwidth]{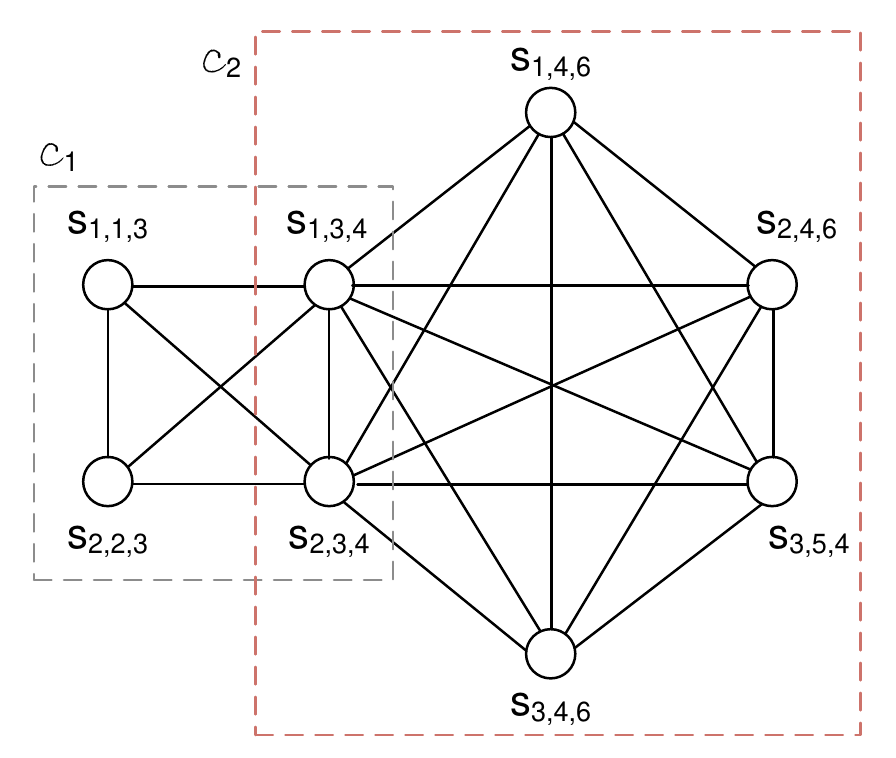}	 
		\caption{Conflict graph corresponding to the topology in Fig.~\ref{fig:simpletopo}. Each vertex corresponds to a segment of a flow $k$ between base stations $i$ and $j$. Cliques highlighted with dashed lines.}
		\label{fig:conflict_graph}
	\end{center}
	\vspace*{-1em}
\end{figure}

\noindent We note that a flow segment can belong to multiple cliques and denote $\cal{C}$ the set of all cliques. We exemplify the conflict graph and clique notions with the simple topology depicted in Fig.~\ref{fig:simpletopo}, for which we can construct the equivalent conflict graph shown in Fig.~\ref{fig:conflict_graph}. Observe that two cliques exist in this example and the segments of flows 1 and 2 over link $l_{3,4}$, i.e. $s_{1,3,4}$ and $s_{2,3,4}$, simultaneously belong to both.\footnote{In this example, cliques are only formed as a results of single-transceivers operating at each node and no secondary interference can be observed. Had node 1 been on the same direction as the (4,6) link, $s_{1,1,3}$ would have formed a third clique with $s_{1,4,5}, s_{2,4,6},$ and $s_{3,4,6}$.}
Returning to our problem, by (\ref{eq:clique_constraint}) we introduce a \emph{clique constraint} that guarantees the total time consumed by all flow segments in a clique does not exceed~1-$\tau$, where $\tau$ is the fraction of time consumed with beamform training operations, i.e. $\sum_{s_{k,i,j} \in C_q}  t_{k,i,j} \leq 1-\tau, \forall k \in {\cal{F_A}}, \forall C_q \in \cal{C}.$

In solving our problem, it will also prove useful to work with the notion of \emph{conflict node}, defined on the actual network topology as below.

\begin{definition}
 In a backhaul network, a `conflict node' is a base station that forwards traffic on behalf of others.
\end{definition}

\noindent For the example shown in Fig.~\ref{fig:simpletopo}, base stations 3 and 4 are conflict nodes.

\subsection*{Solution Existence}

To verify whether a solution to the problem (\ref{eq:objective})--(\ref{eq:clique_constraint}) exists, i.e. max-min fair allocation in a multi-hop mm-wave network is feasible, we first characterise the network's rate region.

\begin{lemma}
\label{lm:rate_region}
The rate region of a multi-hop mm-wave backhaul network is convex.

\begin{proof}
Since we consider transmissions between base stations are precisely scheduled, channel access in a clique can be seen as a single-hop time division multiplex (TDM) instance, which is known to have a convex capacity region~\cite{Clover:2006}. 
The throughput of any sub-flow $s_{k,i,j}$ in a clique $C_q$ is upper bounded by the minimum between the throughput allocated in the clique $C_{q-1}$ traversed previously and the total flow demand $d_k$. The network rate region is obtained by the appropriate intersection of the rate regions of the component cliques. Thus it is convex.
\end{proof}
\end{lemma}

The following key result follows.
\begin{corollary}
 Max-min fair allocation in multi-hop mm-wave networks exists and it is unique.
\begin{proof}
 We can prove by contradiction following the approach of Radunovic and Le Boudec that a max-min fair allocation vector is achievable on compact convex sets~\cite{Radunovic:2007}. As per Lemma~\ref{lm:rate_region} above, the rate region of a scheduled mm-wave backhaul is convex, therefore a max-mix fair rate allocation vector exists. By Theorem 2 in~\cite{Jaffe:81} and the constructive proof of Gafni and Bertsekas, p.1011 in \cite{gafni84}, if any max-min allocation vector exists, then it is unique. 
 
 Hence, in the mm-wave backhaul scenario we consider, a max-min fair rate allocation vector exists and it is unique.
\end{proof}
\end{corollary}

Finally, the rate region has the free disposal property~\cite{Radunovic:2007} since each element of the rate vector $\mathbf{r} = \{r_k\ |\ k \in \cal{F} \}$ is lower bounded by zero and any non-zero feasible allocation can always be decreased. It follows that a progressive filling algorithm can be employed to find the solution to the max-min fair allocation problem with mm-wave particularities. 

\section{WiHaul: Max-min Fair Backhauling}

In what follows we present a max-min fair multi-hop mm-wave backhauling mechanism, which we name {\sc WiHaul}. This consists of a progressive filling algorithm that solves the optimisation problem (\ref{eq:objective})--(\ref{eq:clique_constraint}) in polynomial time, and a light-weight scheduling protocol that distributes airtime solutions among base stations, ensuring they communicate at the right time for the computed durations. Our solution handles mm-wave specific PHY impairments such as link blockage, as the progressive filling routine updates airtime allocation as a result of changes in the rate regions when such events occur. The scheduling procedure further handles terminal deafness (and secondary interference), as it builds on the notion of clique introduced above, which ensures appropriate spatial reuse while  transceivers and receivers always have their beams aligned when intending to communicate.

\begin{algorithm}[t]
\small
\caption{Progressive Filling}
\label{alg:pseudoWF}
\begin{algorithmic}[1]
        \State{$r_k = 0, \forall k$} \Comment{Initialisation}\label{ln:init_rate}
	\State{$\mathcal{F_A} : = \cal{F}$} \label{ln:init_set}
	\Comment{Set of active flows}
	\While{$\mathcal{F_A} \neq \emptyset$} \Comment{Loop until all flows allocated}\label{ln:loop}
		\State{$r_{k} += \epsilon, \forall f_k \in \mathcal{F_A}$} \Comment{Increase rates of all active}\label{ln:increase}
		\Statex{\hfill flows with same step}
		\For{$\forall f_k \in \mathcal{F_A}$}	
			\If{$r_{k} \geq d_k$} \Comment{Flow demand satisfied}\label{ln:demand}
			\State{$r_k := d_k$};
			\State{$\mathcal{F_A} = \mathcal{F_A} \setminus \{f_k\}$} \Comment{Remove flow from active set}\label{ln:remove}
			\EndIf
		\EndFor
		\For{$q = 1: |\cal{C}| $}	\Comment{Loop over all cliques}
		    \State{$t_{k,i,j} = r_{k} / c_{i,j},\  \forall s_{k,i,j} \in C_q$}  \Comment{Time consumed by}
		    \Statex{\hfill each flow segment in $C_q$}
		    \If{$\sum_{s_{k,i,j} \in C_q} t_{k,i,j} \geq 1$}   \Comment{Clique constr. not met}\label{ln:clique}
			\State{$t_{\text{left}} = 1$} \Comment{Total airtime budget}\label{ln:totalair}
			\State{$S = 0$} \Comment{Sum of inverse capacities of}
			\Statex{\hfill links traversed by active flows}
			\For{$\forall s_{k,i,j} \in C_q$} \Comment{Loop over all sub-flows}
			  \If{$f_k \in {\cal{F}} \setminus {\mathcal{F_A}} $} \Comment{Flow inactive}
			    \State{$t_{\text{left}} = t_{\text{left}} - t_{i,j,k}$} \Comment{Subtract airtime}\label{ln:reserved}
			    \Statex{\hfill already reserved}
			  \Else \Comment{Flow active}
			    \State{$S = S + 1/c_{i,j}$} \Comment{Update sum for} \label{ln:sum}
			    \Statex{\hfill subsequent airtime weighting}
			  \EndIf
			\EndFor
			  \State{$R = t_{\text{left}} / S$} \Comment{Rate to allocate for all active flows}\label{ln:new_rate}
			  \For{$\forall f_k \in \mathcal{F_A}$} \Comment{Loop over all active flows}
			    \State{$r_{k} = R;\ t_{k,i,j} = r_{k} / c_{i,j}$} \Comment{Allocate rate and}\label{ln:airtime}
			    \Statex{\hfill airtime on each link}
			    \State{Freeze $r_k;\ \mathcal{F_A} = \mathcal{F_A} \setminus \{f_k\}$}\label{ln:remove2} \Comment{Freeze rate}
			    \Statex{\hfill remove flow from active set}
			  \EndFor
		    \EndIf
	    \EndFor
	\EndWhile
\end{algorithmic}
\end{algorithm}

\subsection{Progressive Filling Algorithm}
\label{sec:progressive_filling}

Algorithm \ref{alg:pseudoWF} summarises the progressive filling procedure we propose to achieve max-min fair allocation of the backhaul resources under clique and demand constraints, and we detail its operation next. 
We start with all flow rates equal to zero and consider none of the aggregate flows have been allocated resources (lines~\ref{ln:init_rate}--\ref{ln:init_set}). We call active flows, those which have not been allocated resources yet. We gradually increase flow rates simultaneously, in steps of size $\epsilon$ Kbps (line~\ref{ln:increase}) until one or more flows either meet their demands (line~\ref{ln:demand}) or activate a clique constraint (line~\ref{ln:clique}). Note that $\epsilon$ is a configurable parameter whose magnitude impacts on algorithm runtime. If a flow's demand $d_k$ is satisfied, we freeze the allocated rate $r_k$ to the demand and remove that flow from the active set (line~\ref{ln:remove}), thereafter considering it inactive and its resources frozen.

When a clique is fully utilised, we stop increasing the rates of the flows traversing it and proceed with computing these rates from scratch. To this end, we subtract from the total available airtime, i.e. 1 (assuming beamform training $\tau << 1$, on average, line \ref{ln:totalair}), the fractions already reserved for \emph{inactive flows} (line~\ref{ln:reserved}) and sum up the inverse of the link capacities corresponding to \emph{active flows} in that clique (line~\ref{ln:sum}). The latter will allow us to provide all active flows with the same rate $R$ (line~\ref{ln:new_rate}), which under heterogeneous link rate conditions translates into allocating airtimes to each sub-flow that are inversely proportional to the traversed link's capacity (line~\ref{ln:airtime}), i.e.
\begin{align*}
 t_{k,i,j} = \cfrac{t_\text{left}}{ c_{i,j} \mathlarger{\mathlarger{\mathlarger{\sum}}}\limits_{s_{k,l,m} \in \mathcal{F_A} \cap C_q}\ \cfrac{1}{c_{l,m}}}.
\end{align*}

\noindent It is straightforward to verify that airtimes $t_{k,i,j}$ above sum to $t_\text{left}$, as required. Subsequently, we freeze the rates $r_k$ of flows in clique $C_q$ and remove them from the active set (line~\ref{ln:remove2}).

We repeat this procedure for the remaining active flows, until meeting their demand or activating other clique constrains. The progressive filling algorithm terminates when the set of active flows is empty (line~\ref{ln:loop}). At that point we have obtained the airtimes to be allocated for each flow on each traversed backhaul link, in order to fulfil the max-min fair allocation of the rates. 

Our algorithm's runtime is a function of the highest flow rate divided by the step-length, which recall is configurable, and the total number of flows. Therefore the algorithm solves the max-min fairness optimisation problem posed in polynomial time. The results we present in Sec.~\ref{sec:runtime} confirm this assessment.

\subsection{Scheduling Procedure}
\label{sec:wihaulprotocol}
Terminal deafness is a major challenge in mm-wave networks. Therefore, unless stations know to which neighbour to steer their beams, when, and for how long, they may be locked out, which would lead to frame loss and overall performance degradation. Such degradation may also occur when beams of different communicating pairs partially overlap, resulting in secondary interference. Algorithm \ref{alg:pseudoWF} described previously addresses the computation of airtimes for each flow segment, in order to attain max-min fair rates. To convey the computed airtimes and overcome TX/RX issues, i.e. deafness or secondary interference, {\sc WiHaul} employs a network-wide co-ordination procedure based on a scheduling hierarchy. This enables a centralised controller to dictate when nodes can transmit to others without conflict and in which order, so as to maximise spatial reuse. This effectively means that scheduling will also circumvent any potential terminal deafness.

Algorithm~\ref{alg:scheduling} gives the pseudocode of {\sc WiHaul}'s scheduling operation, which we explain next with the example topology shown in Fig.~\ref{fig:simpletopo}. We assume a central controller (typically placed at the gateway; here node~6) has full knowledge of the network topology, including the hop distance to each base station, which of these are conflict nodes (i.e. have more than one neighbour), as well as their addresses, i.e.
\begin{enumerate}
  \item[(1)] $H_i$: hop distance from node $i$ to the gateway,
  \item[(2)] $S_i$: node $i$'s conflict state,     
   \begin{equation*}
  S_i = 
  \begin{cases} 1, & \text{if $i$ is a conflict node,}\\
  0, & \text{if $i$ is a leaf node;}
  \end{cases} 
 \end{equation*}
  \item[(3)] $A_i$: node $i$'s unique ID (e.g. its IP address).
\end{enumerate}
\vspace*{0.5em}

\begin{figure}[t]
	\begin{center}
		\includegraphics[width=1\columnwidth]{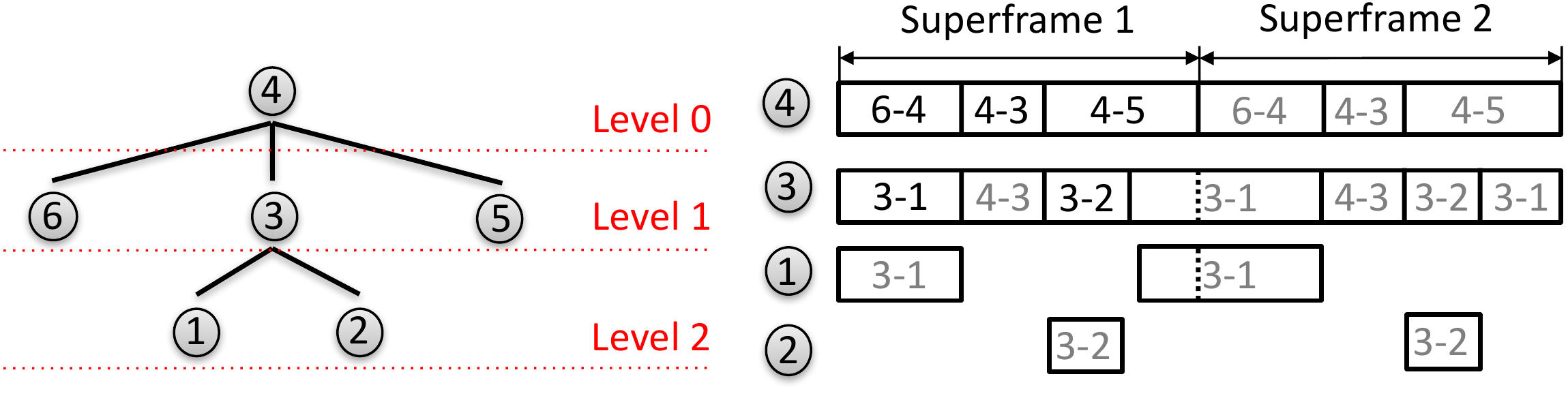}
		\caption{Hierarchical scheduling structure corresponding to the topology in Fig.~\ref{fig:simpletopo} (left) and time slots allocated for each communicating pair transporting different flow segments (right). Links $l_{4,5}$ and $l_{3,1}$, and respectively $l_{6,4}$ and $l_{3,2}$ can be simultaneously active. Time slots labelled in black represent those being scheduled by node circled; those in grey time slots by the parent of current node.}
		\label{fig:hierarchy}
		\vspace*{-1em}
	\end{center}
\end{figure}

\begin{algorithm}[t]
\small
\caption{Max-min Fair Scheduling}
\label{alg:scheduling}
\begin{algorithmic}[1]
\State{Obtain air time shares $t_{k,i,j}, \forall k,i,j$ with Algorithm~\ref{alg:pseudoWF}} 
\State{${\cal H} = $ \Call{Build scheduling hierarchy}{network topology}}\label{ln:hierarchy}
\State{Root coordinator of ${\cal H}$ assigns slots to its child nodes, i.e. Level 1 nodes, given total airtime available} 
\While{!bottom level of ${\cal H}$}\label{ln:loop2}
    \State{Order conflict nodes by $A_i$ in increasing order}
	\ForAll{conflict nodes}
	    \State{Accept airtime assigned by parent node}\label{ln:accept}
       \ForAll{child nodes of current parent}
                \If{node's priority lower than others in clique} 
                    \State{Mark time slots used by other nodes as taken} 
                \EndIf
            \State{Assign airtime to child nodes}\label{ln:assign}
        \EndFor
        \State{Move to the next level}
    \EndFor    
\EndWhile
\Statex
\Function{Build scheduling hierarchy}{topology}
  \State $\cal{L} \leftarrow$ 0; \Comment Level 0
  \State Set node with $H_c = \min_{\{i | S_i =1\}} H_i$ as root coordinator
  \State Place the root coordinator on $\cal{L}$
  \While {!(all nodes assigned a level)}
    \State ${\cal{L} \leftarrow \cal{L}} + 1$ \Comment{Advance level}
  	\State {Place on current level nodes $i$ with $|H_i - H_c| = \cal{L}$}
  	\label{ln:place-node} 
  	\EndWhile
	\State\Return Scheduling hierarchy ${\cal H}$
\EndFunction
\end{algorithmic}
\end{algorithm}
\noindent With this information and the airtime shares computed by Algorithm~\ref{alg:pseudoWF}, the controller constructs a hierarchy to establish when a node should transmit/receive and when it should schedule its neighbours, respectively (line~\ref{ln:hierarchy}). Specifically, {\sc WiHaul} first considers all conflict nodes as eligible candidates for acting as scheduling coordinators (in our example nodes~4 and 3). Among these, the one with the lowest hop distance $H_c = \min_{\{i | S_i =1\}} H_i$ is designated as the root coordinator and placed at the top of the scheduling hierarchy, namely at Level 0. In this example it is node~4 that acts as coordinator, while 6 (the gateway) is not a conflict node. The remaining nodes with $S_i = 1$ will be placed at a level that depends on the difference between their $H_i$ value and that of the main coordinator ($H_c$) i.e. Level $i = |H_i - H_c|$ (line~\ref{ln:place-node}). Nodes with $S_i = 0 $ will be placed at Level$_i$ below their neighbouring conflict node. As such, in our example nodes 5 and 6 reside at Level 1, while 1 and 2 are placed at Level 2, as illustrated in Fig.~\ref{fig:hierarchy}.

At each level of the hierarchy, {\sc WiHaul} assigns airtime top--down, a node accepting the time allocated by its parent and assigning airtime to its children (lines~\ref{ln:accept}--\ref{ln:assign}). In the considered example, the protocol first assigns time for 4 and then the nodes at Level 1, i.e. 3, 5 and 6. In turn, node 3 assigns time to 1 and~2, outside the interval when it is involved in communication with 4. This allows for spatial reuse, as links $l_{4,5}$ and $l_{3,1}$, and respectively $l_{6,4}$ and $l_{3,2}$ will be active simultaneously.

In case of multi-path routing, it may happen that two or more nodes on the same level share the same neighbouring node that they could schedule. In such cases, the node with the smallest identifier $A_i$ takes priority and will be the one scheduling. In turn, the child informs the other candidate parents of the assigned time, to resolve the tie and avoid conflicts. This process is repeated until all computed airtimes have been disseminated to all stations. 

Subsequently, nodes will periodically switch their beams towards the corresponding neighbours for transmission/reception during the assigned times. To adapt to the dynamics of physical channel conditions (e.g. link blockage) and the changing flow demands, the controller will periodically (e.g. every superframe) collect link quality and flow demand information, run the progressive filling algorithm, and re-schedule flow segments as appropriate.  

\section{Performance Evaluation}
\label{sec:simulationevaluation}

To evaluate the performance of {\sc WiHaul}, we implement this in NS-3 and conduct extensive simulations under different scenarios,\footnote{The source code of our implementation is available at {\ttfamily \url{https://git.io/wihaul}}.} comparing with recent scheduling schemes for mm-wave networks, including DLMAC~\cite{sim2016learning},  MDMAC~\cite{singh2010distributed}, and variations of these. We examine achievable gains in terms of flow throughput distribution and overall network throughput, and the level of fairness each approach attains over realistic multi-hop topologies. We further analyse {\sc WiHaul}'s behaviour in terms of allocated flow throughputs and airtimes, 
and give insight into the impact of link rates and flow demands on the partitioning of resources.
Lastly, we evaluate our solution with real data traffic traces and examine end-to-end delay performance. 

It is worth noting that making a definite comparison of the complexity entailed by our solution and the benchmarks considered is difficult. This is largely due to the different paradigms employed, i.e. centralised vs distributed, and random vs scheduled channel access. Unlike our proposal, the benchmarks are also subject to convergence times that depend on neighbourhood size and payload lengths, and may require restarting to cope with traffic dynamics. Slot alignment is also problematic in distributed settings, yet not explicitly discussed by the respective authors. In our case, the airtime allocation is a function of the highest flow rate and a configurable step length, while scheduling runtime depends on the number of nodes in a given topology.

\subsection{Simulation Environment}
\label{sec:simulationSetup}

While our solution is applicable to any multi-hop mm-wave backhauls that operate in a scheduled mode, for evaluation purposes we employ the 802.11ad PHY with the SP based MAC, as this standard is already mature.
To incorporate multi-hop frame relaying, the controller logic, and the progressive filling algorithm in NS-3, we extended the IEEE 802.11ad SP based MAC implementation of Facchi \emph{et al.}~\cite{facchi:2017}. 
The simulator incorporates the 802.11ad MAC frame structure and simple PHY functionality for directional multi-gigabit OFDM transmissions. The Beacon Interval (BI) header occupies a configurable fraction of the BI duration. We use the default IEEE 802.11ad setting, i.e. 10\% of the BI duration for overhead, which is also in agreement with the LTE and 5G NR frame structures.\footnote{LTE and subsequent 5G NR Type 2 frames for TDD access dedicate 10\% of the frame duration for Downlink Pilot Time Slot (DwPTS), Guard Period (GP), and Uplink Pilot Time Slot (UpPTS), to handle TDD operation specifics.\cite{3GPP-NR}} Overall, this overhead interval is reserved for beamform training, control message exchange, and schedule dissemination. Specifically, should there be any changes in link capacity, routing, or flow demand, as we will discuss in details in Sec.~\ref{sec:dynam}, the BI overhead will cover the time required to propagate to backhaul nodes the flow rate allocations re-computed by the controller. Actual packets are exchanged during the data transmission interval (DTI), as scheduled by {\sc WiHaul}. We employ the Friis path loss model based on which the received power and the SNR are computed. We then map the SNR to a specific modulation and coding scheme (MCS), which corresponds to the link capacity, $c_{i,j},~\forall \{i,j\}$. 
Given the switched operation of transmissions and receptions, and the high PHY bit rates employed on links, to avoid excessive delays and buffer overflows at relaying stations, we divide the airtime allotted to each sub-flow into multiple SPs, each of shorter duration. In the simulation evaluation, we work with 20 short SPs that sum up to the computed airtime allocations. 

We implement a central controller that executes the proposed {\sc WiHaul}, including progressive filling and scheduling, and incorporate measurements of MAC queue length to monitor events such as buffer overflows. 
If changes in flow demand or link capacity take place, the progressive filling and scheduling operations will be triggered to perform allocations for the next BI.
Further, the simulation tool incorporates MAC protocol data unit aggregation (A-MPDU) and MAC frame relaying, to support efficient multi-hop backhauling scenarios. We assume that during DTI, the beams of TX/RX base stations are perfectly aligned. While PHY design remains outside the scope of this work, we investigate the impact of secondary interference and show how {\sc WiHaul} can tackle such issues, in Sec.~\ref{sec:2ndintf}

The NS-3 build-in module allows for full-stack simulation including application, transport and internet layers on top of the 802.11ad MAC and PHY. We work with applications that generate fixed packets of 1470 Bytes, except when experimenting with real traffic traces.
We summarise the parameters used in simulation in Table~\ref{tab:settings}.

\begin{table} [htb]
\centering
\begin{tabular}{ l l }
\toprule
\textbf{Parameter} &Value\\\hline \hline
Tx power & 10dBm\\ 
Tx/Rx antenna gain & 20dB\\ 
BI duration& 102,400$\mu$s \\ 
BI overhead& 10,240$\mu$s\\
Progressive filling step length ($\epsilon$) & 10kbps\\ 
UDP payload& 1,470B\\ 
TCP MSS& 1,460B\\
TCP Initial Slow Start Threshold& 64KB\\
TCP Tx/Rx Buffer Size& 10MB\\\bottomrule
\end{tabular}
\caption{Simulation settings.}
\label{tab:settings}
\end{table}
\vspace*{-1em}

\subsection{Fairness Metrics}
Note that max-min is a \emph{qualitative} fairness criterion. That is, some allocation is max-min fair if increasing the rate of a flow is only possible by decreasing that of others~\cite{Bertsekas:1992}. Unlike e.g. Jain's fairness index, this typically does not have a directly measurable value. Therefore, to quantify fairness, we first resort to the concept of inequality distribution used in economics, and compute Gini coefficients \cite{gini:1921}, using the following formula:
\[
 \text{G} = \frac{\sum_{k=1}^{n}\sum_{l=1}^{n}|r_k -  r_l| }{2n\sum_{k=1}^{n}r_k},
\]
where $r_k$ is the rate allocated to flow $k$, and $n$ is the total number of flows. The lower this coefficient is with a certain rate allocation vector, the more fair the distribution of resources is.

To add further perspective and quantify to what extent the minimum flow rate in the network might be higher with {\sc WiHaul} than with other schemes, we employ the generalised measure of fairness defined in \cite{lan:2010}, as follows 
\[
 \mathcal{M}_{\beta}(\mathbf{r}) = \text{sign} (1-\beta) \cdot \Bigg [\sum_{k=1}^{n} \Bigg(\frac{r_k}{\sum_{l}r_l}\Bigg)^{1-\beta}\Bigg] ^{\frac{1}{\beta}},
\]
where $\beta$ dictates different types of fairness measures. For max-min fairness $\beta \rightarrow \infty$, and $\mathcal{M}_{\beta}(\mathbf{r})$ becomes
\begin{align*}
 \mathcal{M}_{\beta}(\mathbf{r}) &= \lim_{\beta\rightarrow\infty} \sign(1-\beta) \left[\sum_{k=1}^n \left(\frac{r_k}{\sum_{l=1}^n r_l}\right)^{1-\beta}\right]^{\frac{1}{\beta}}  
\end{align*}
\begin{align*}
  & = - e^{\lim_{\beta\rightarrow\infty} \log \left[\sum_{k=1}^n \left(\frac{r_k}{\sum_{l=1}^n r_l}\right)^{1-\beta}\right]^{\frac{1}{\beta}}}. 
\end{align*}

We denote $y_k = (\sum_l r_l)/r_k$ and solve the limit above by applying l'H\^opital's rule, which leads to
$
 \lim_{\beta\rightarrow\infty} (\sum_{k=1}^n y_k^{\beta-1} \log(y_k))/(\sum_{k=1}^n y_k^{\beta-1}).
$
As $\beta\rightarrow\infty$,~the numerator is dominated by the highest $y_k$ term, i.e. $\max_k \{y_k \log(y_k)\}$, hence the limit converges to $ \max_k { \sum_l r_l / {r_k}}$ and max-min fairness can be measured with

\begin{equation}
 \mathcal{M}_{\beta}(\mathbf{r}) = - \max_k \left\{ \frac{\sum_l r_l}{r_k} \right\}.
 \label{eq:fairnessmeasure}
\end{equation}

\subsection{Comparison with State-of-the-Art Solutions}

\begin{figure*}[!h]
	\begin{center}
		\begin{subfigure}{0.225\textwidth}
			\includegraphics[width=\textwidth]{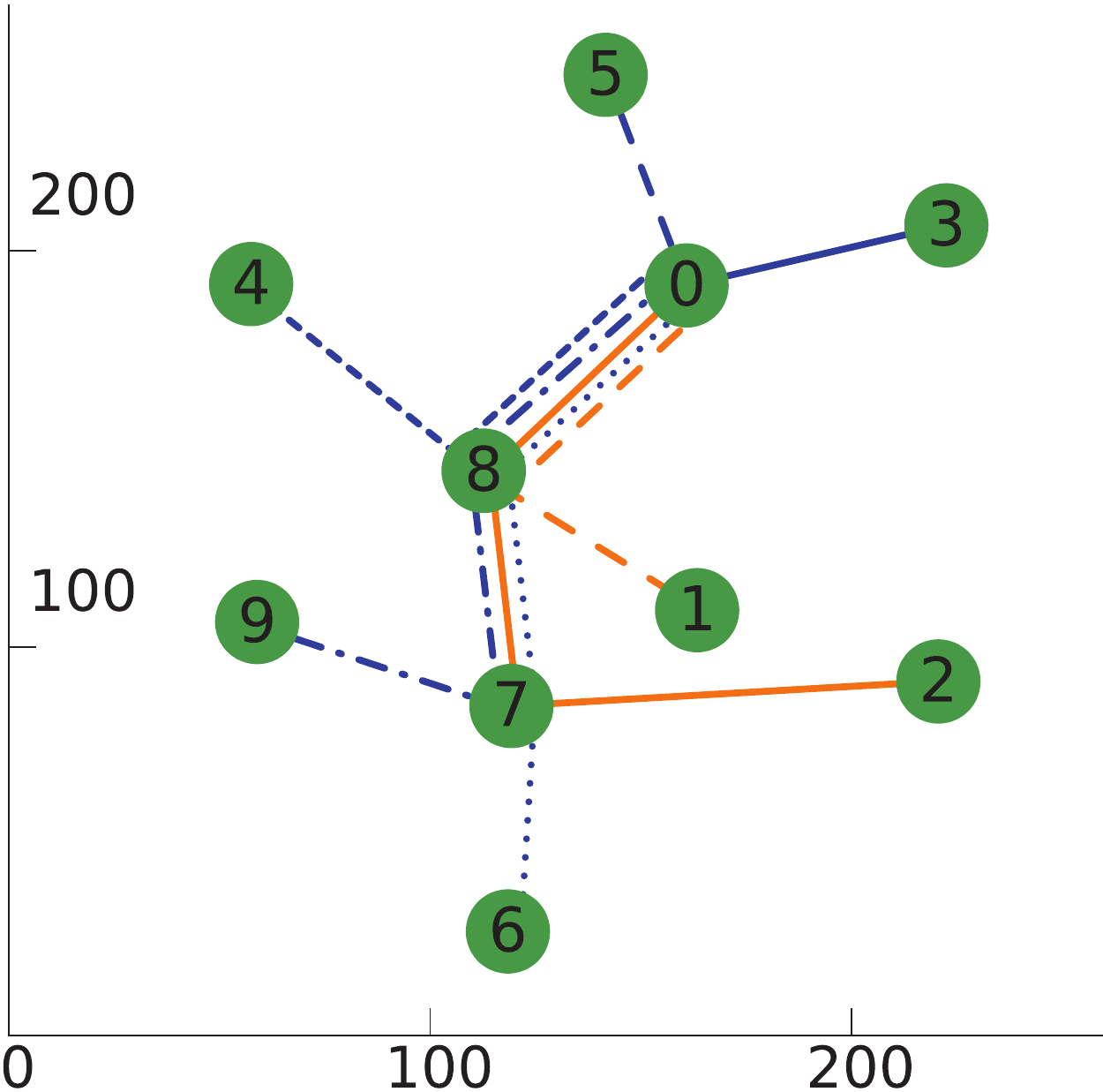}
			\caption{Topology 1.}
			\label{fig:topo1map}
		\end{subfigure}
		\begin{subfigure}{0.225\textwidth}
			\includegraphics[width=\textwidth]{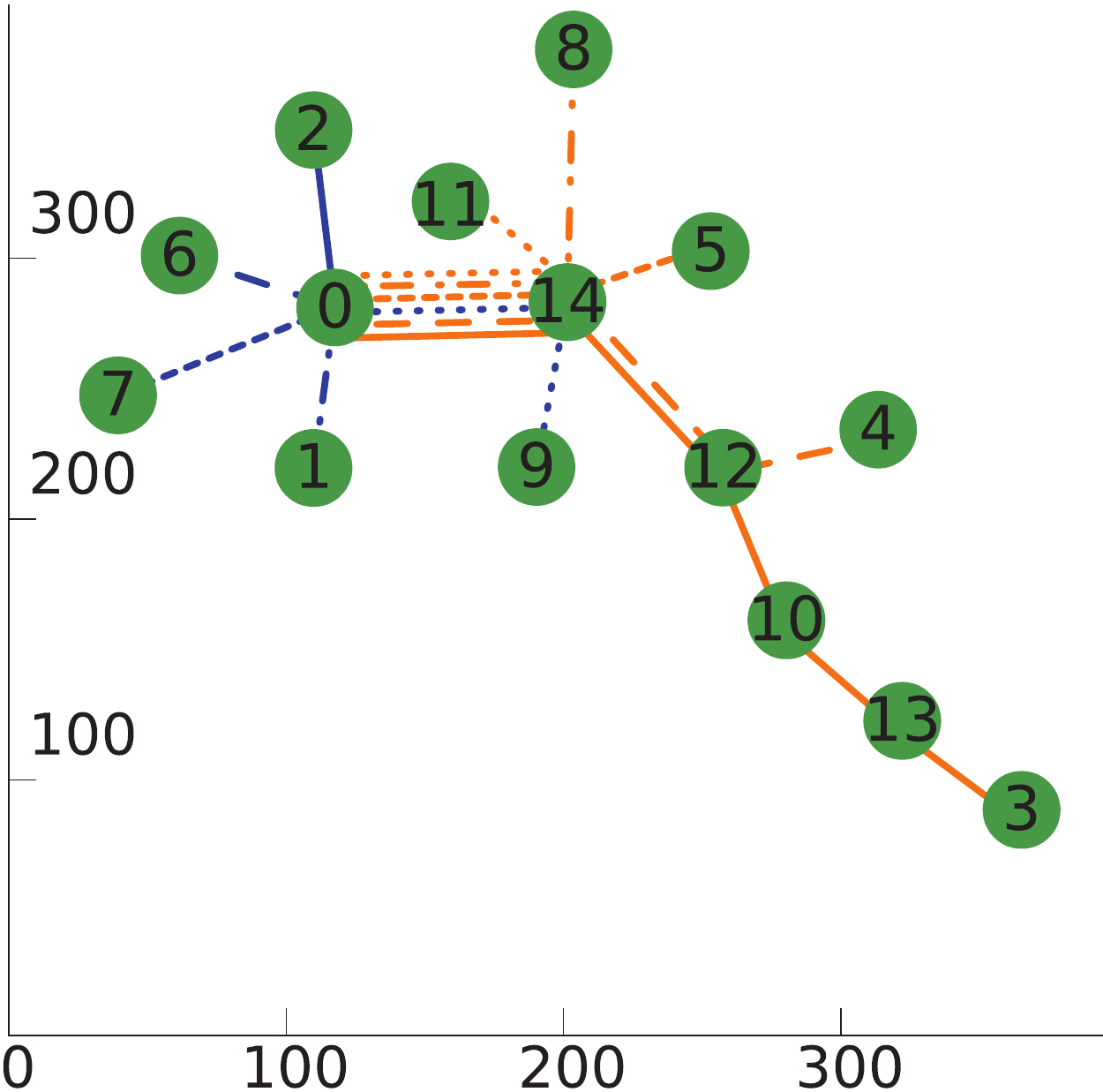}
 			\caption{Topology 2.}
			\label{fig:topo2map}
		\end{subfigure}		
		\begin{subfigure}{0.225\textwidth} 
			\includegraphics[width=\textwidth]{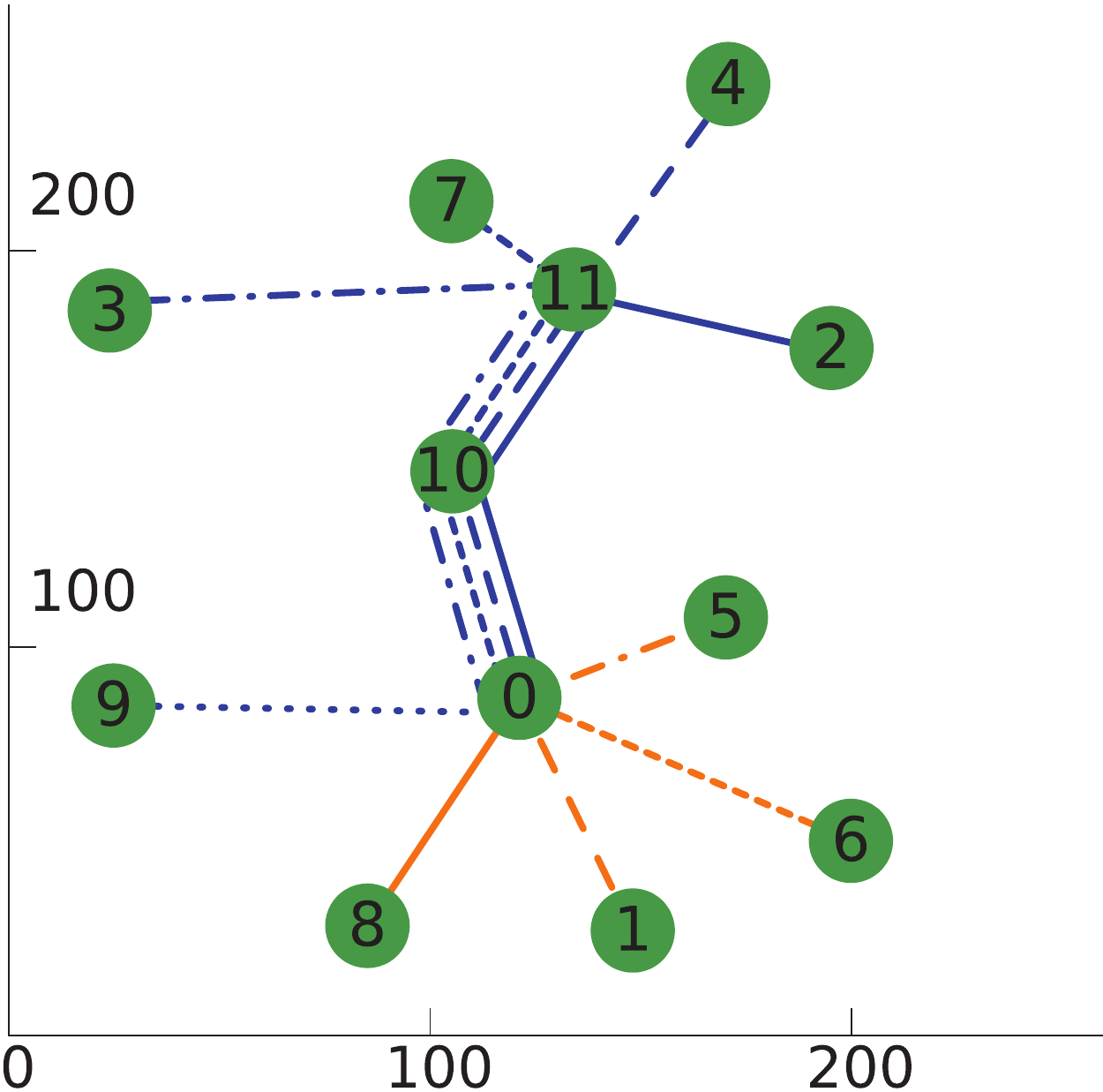}
			\caption{Topology 3.}
			\label{fig:topo3map}
		\end{subfigure}
		\begin{subfigure}{0.225\textwidth}
			\includegraphics[width=\textwidth]{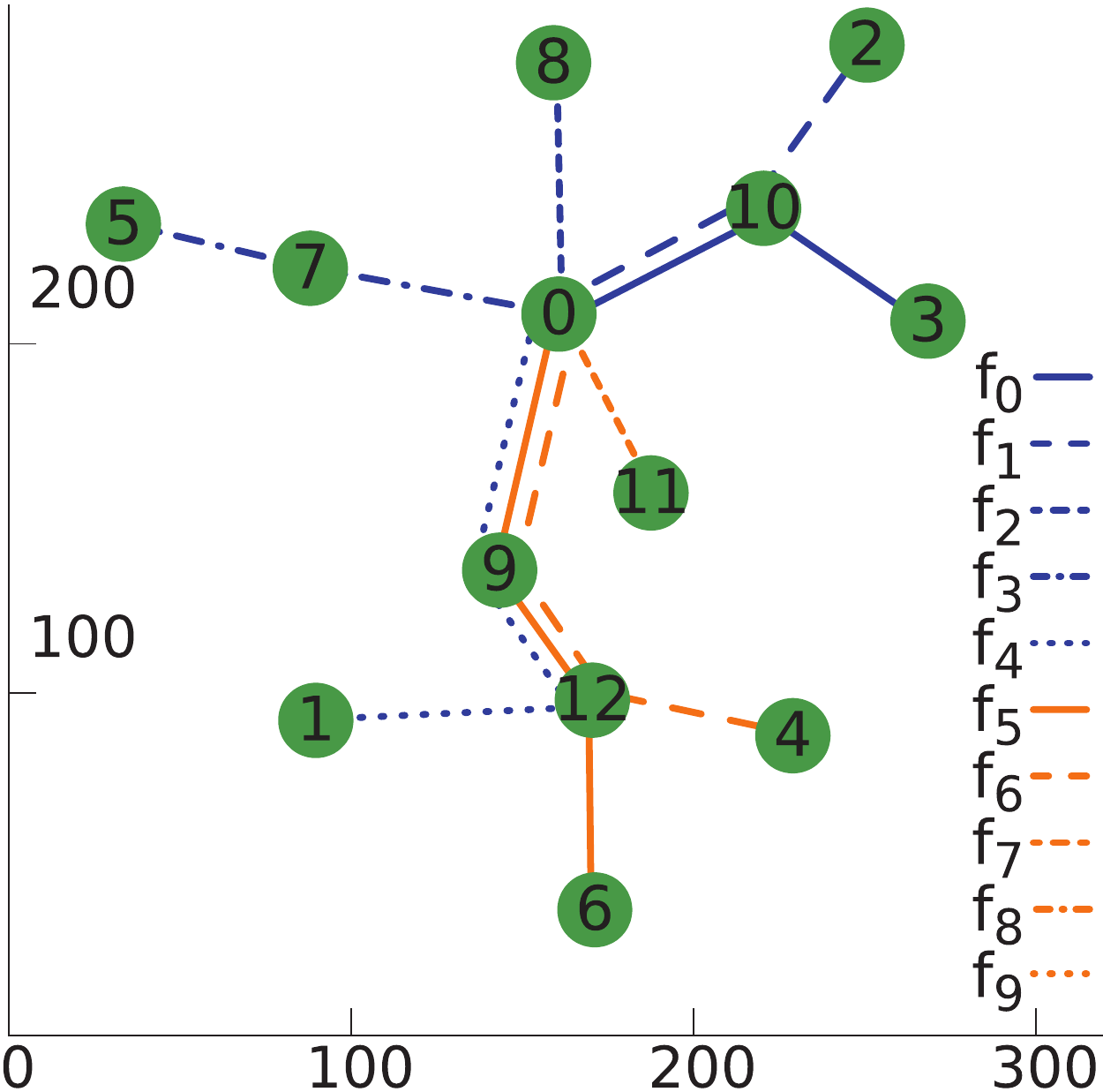}
			\caption{Topology 4.}
			\label{fig:topo4map}
 		\end{subfigure}	
		\caption{Multi-hop topologies used for performance evaluation, generated with the Cerd\`a-Alabern model~\cite{cerda2012topology}.}
		\label{fig:4topos}
		\end{center}
\end{figure*}

We compare the performance of {\sc WiHaul} against that of recent mm-wave scheduling schemes \mbox{DLMAC}~\cite{sim2016learning} and \mbox{MDMAC}~\cite{singh2010distributed} in terms of mean and total network throughput, and inter-flow fairness. We conduct the evaluation over several topologies generated with the Cerd\`a-Alabern model that captures the characteristics of real-world multi-hop wireless deployments~\cite{cerda2012topology}. The topologies considered comprise 10 to 15 stations (including the Internet gateway) 
and the number of aggregate flows traversing the network varies between 7 and 10. We illustrate four of these topologies in Fig.~\ref{fig:4topos}, where the X and Y axes represent the base stations' coordinates, with base station $0$ being the gateway. Link rates vary between 2.772--6.756Gbps, depending on the distance between stations.

We also compare against optimised \mbox{DLMAC} and MDMAC versions that seek to reduce gaps between transmissions (BinDLMAC)~\cite{sim2016learning} and  operate with slot sizes that maximise transmission efficiency respectively (OptMDMAC).\footnote{The default MDMAC design works with a slotted channel where slot size is fixed to 20$\mu$s. The optimised version we consider works with slots that can accommodate exactly one transmission burst.} We note all these are decentralised and do not explicitly consider fairness in their design. 
Each approach transports backlogged aggregate flows 
over UDP. 

\begin{figure*}[!h]
	\begin{center}
		\begin{subfigure}{0.49\textwidth}
			\includegraphics[width=\textwidth]{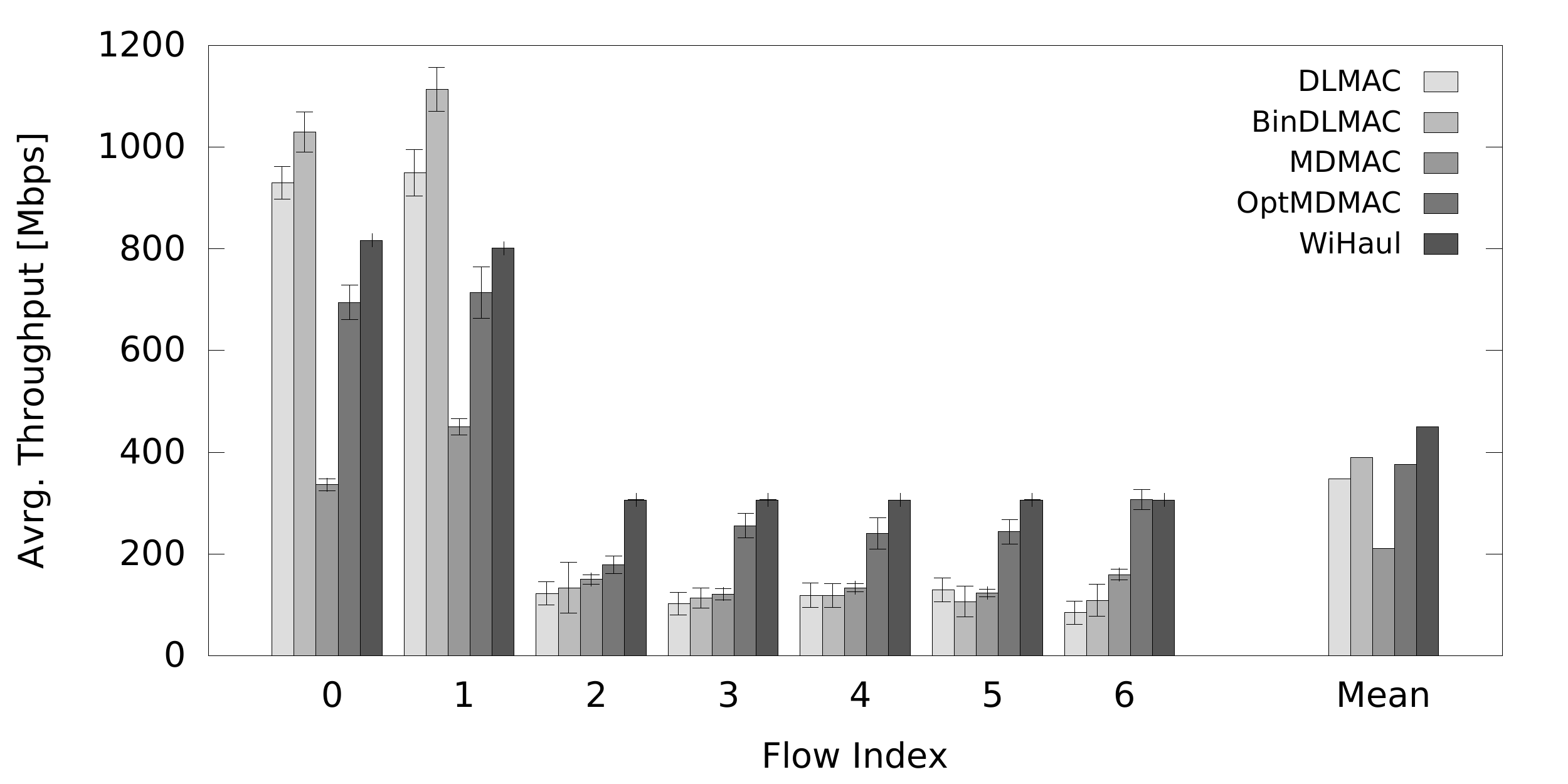}
			\caption{Throughput distribution in Topology 1.}
			\label{fig:topo1thrput}
		\end{subfigure}
		\label{fig:topo6flows}
		\begin{subfigure}{0.49\textwidth}
			\includegraphics[width=\columnwidth]{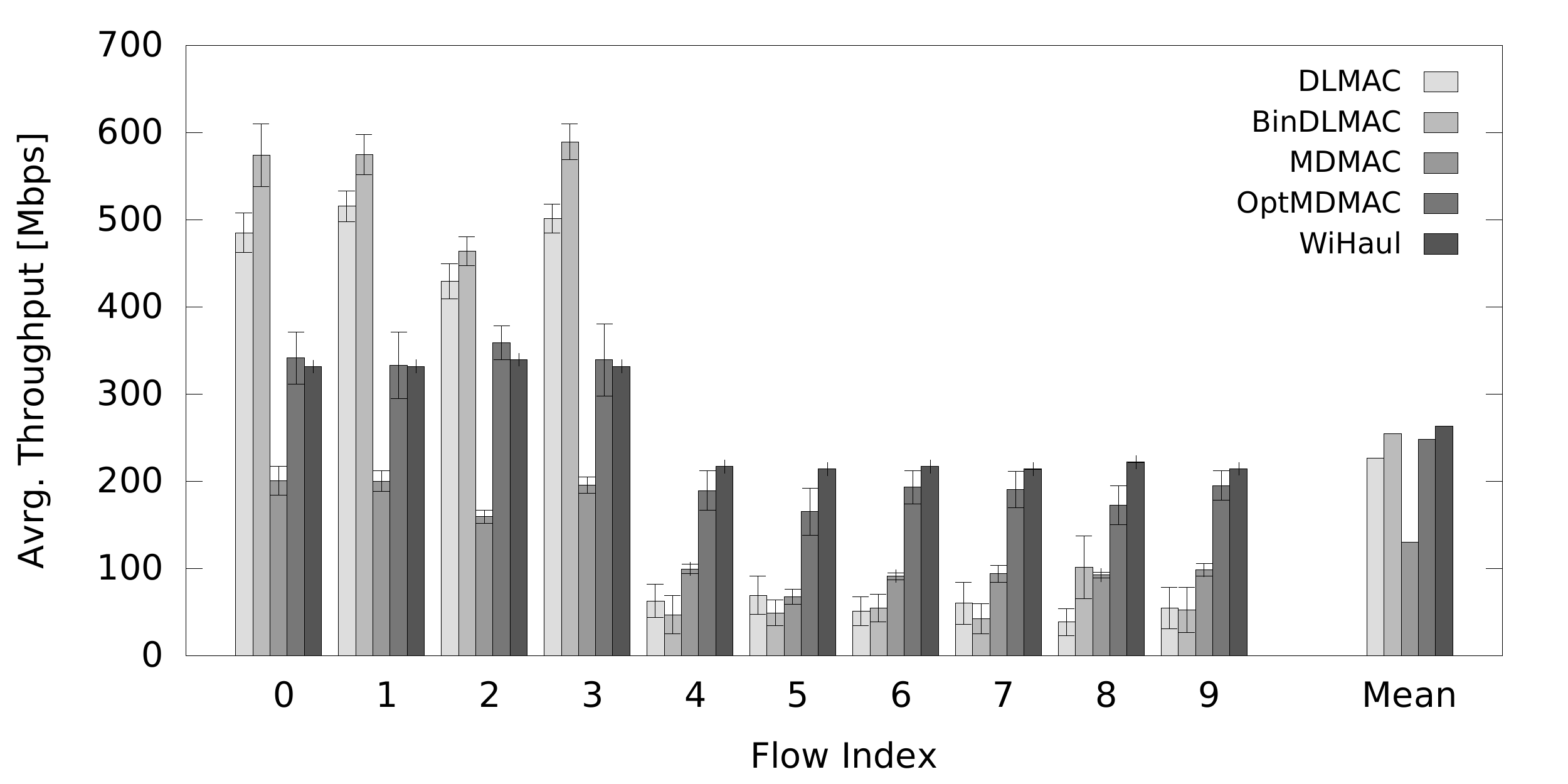}
			\caption{Throughput distribution in Topology 2.}
			\label{fig:topo2thrput}
		\end{subfigure}
		\label{fig:topo9flowssta12} 		
		\begin{subfigure}{0.49\textwidth}
			\includegraphics[width=\columnwidth]{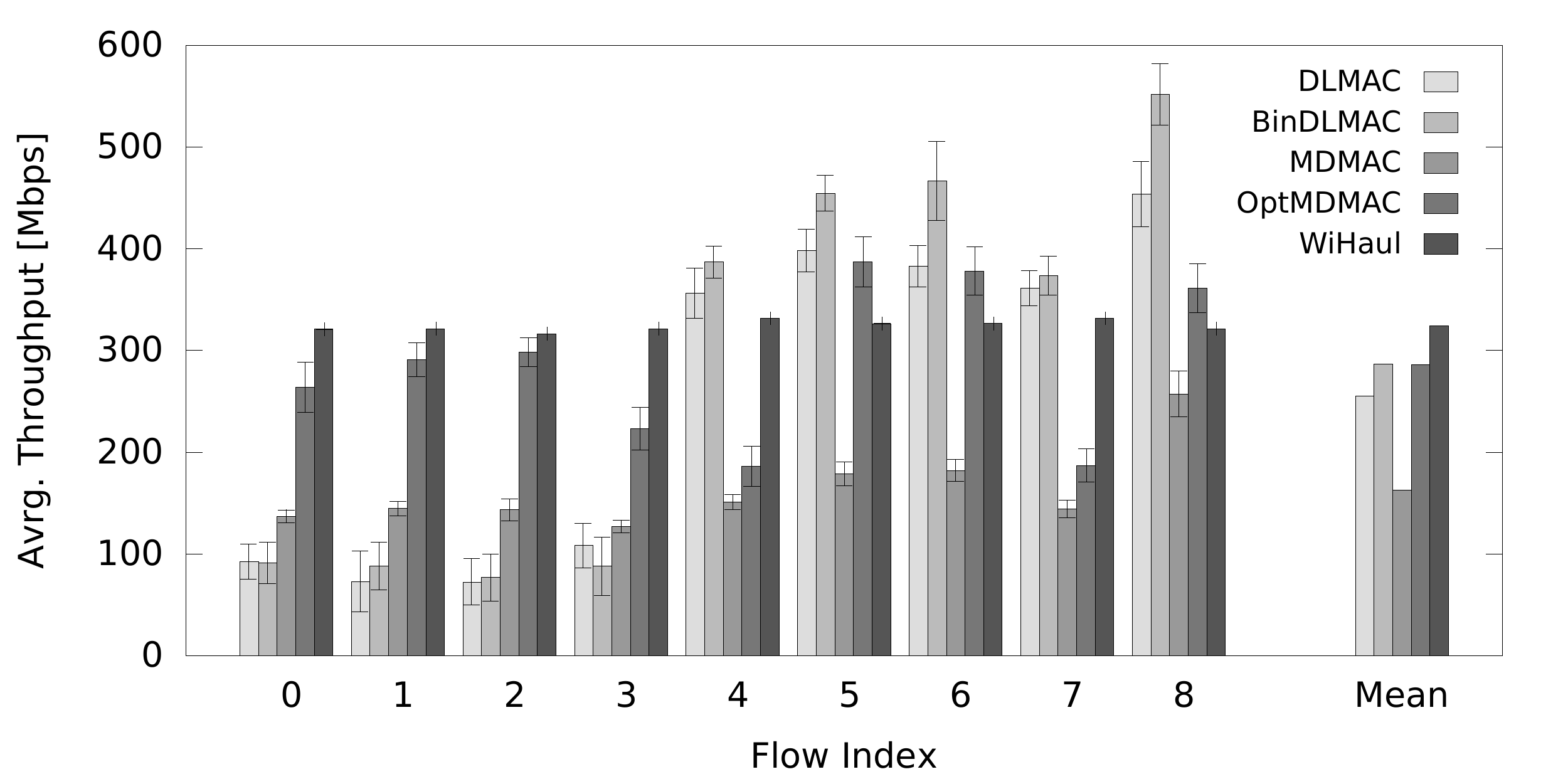}
			\caption{Throughput distribution in Topology 3.}
			\label{fig:topo3thrput}
		\end{subfigure}
		\begin{subfigure}{0.49\textwidth}
			\includegraphics[width=\columnwidth]{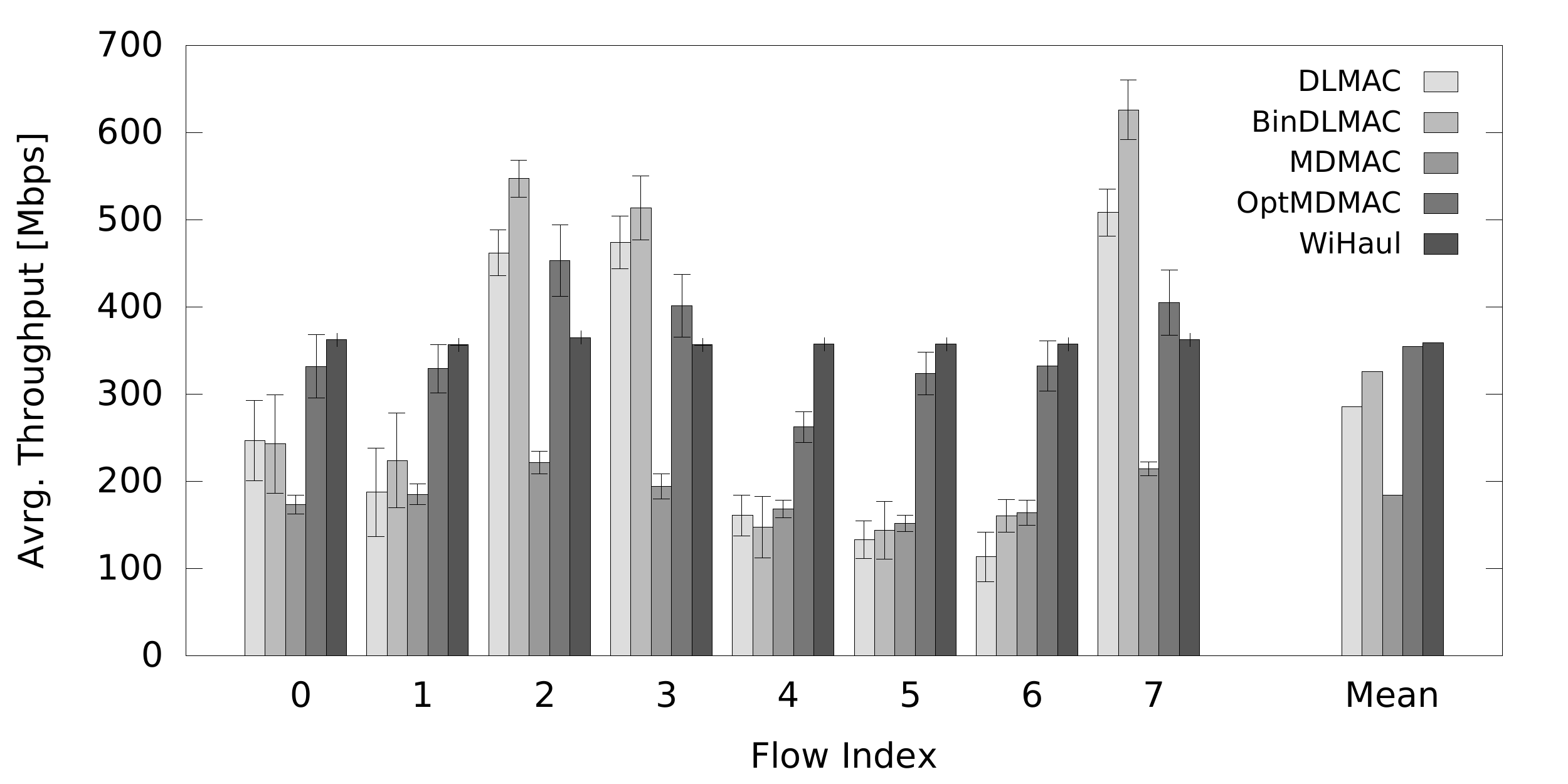}
			\caption{Throughput distribution in Topology 4.}
			\label{fig:topo4thrput}
		\end{subfigure}
		\caption{Throughput comparison of {\sc WiHaul} and existing schemes over the topologies shown in Fig.~\ref{fig:4topos}. Simulation results.}
		\label{fig:thrpt4topos}
	\end{center}
	\vspace{-1em}
\end{figure*}

\begin{figure}
\vspace*{-1.25em}
\includegraphics[width=\columnwidth]{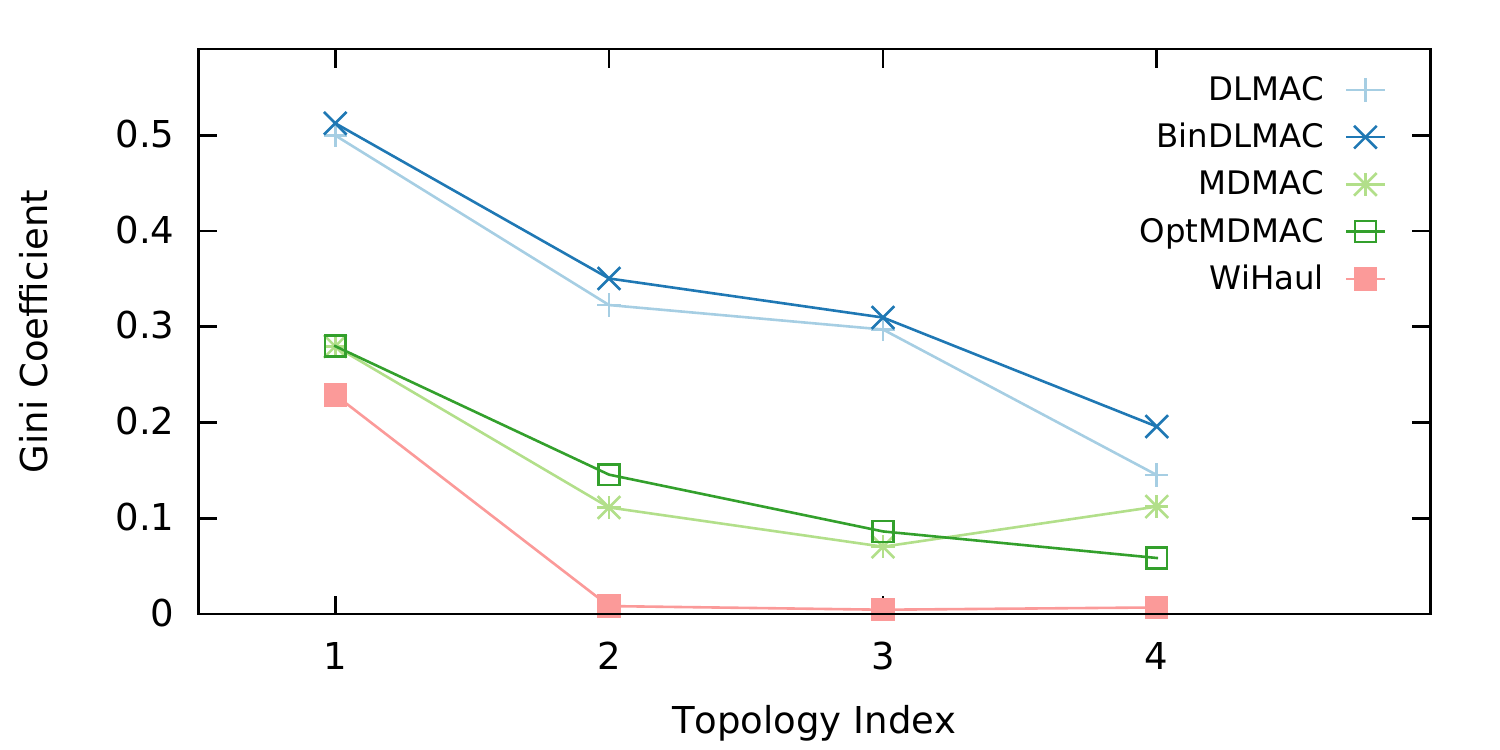}
\caption{Gini coefficients corresponding to the throughput distribution attained by each scheme in topologies in Fig.~\ref{fig:4topos}. Simulation results.}
\label{fig:gini}
\vspace*{-1.5em}
\end{figure}

\vspace*{1em}
\hrule
\vspace*{0.25em}

\noindent\textbf{Finding: {\sc WiHaul} achieves the highest average flow throughput (and therefore total network throughput), irrespective of the number of hops flows traverse and with how many competing flows they share links.} 

\vspace*{0.25em}
\hrule
\vspace*{1em}

Let us examine first Figs.~\ref{fig:topo1thrput}--\ref{fig:topo4thrput}, where we show the average and 95\% confidence intervals of individual flow throughputs attained with {\sc WiHaul}, DLMAC, MDMAC, and their variations, in each topology considered. In these figures we also plot the average throughput performance over all flows as the last cluster of bars to the right of each plot. Observe in these clusters that the bars corresponding to {\sc WiHaul} are indeed the highest and the total network throughput ranges between 2.25-2.5Gbps in all cases.

\vspace*{1em}
\hrule
\vspace*{0.25em}

\noindent\textbf{Finding: With {\sc WiHaul}, flows attain similar throughput as long as they share the same cliques, while additional underutilised network resources are equally divided among unconstrained flows.} 

\vspace*{0.25em}
\hrule
\vspace*{1em}

Indeed, observe that flows which encounter less competition attain superior performance with our approach, without negatively impacting on the others. This can be observed in Figs.~\ref{fig:topo1thrput} and \ref{fig:topo2thrput}, where with {\sc WiHaul} flows $f_0$ and $f_1$, and respectively $f_0$--$f_3$ achieve approximately 450Mbps and 100Mbps more throughput than the other flows traversing the backhaul. At the same time, we reduce the gross performance dissimilarity between flows (e.g. up to 1Gbps between flows $f_1$ and $f_3$ with BinDLMAC in topology 1). In addition, the flows penalised by earlier approaches attain up to 5$\times$ higher throughput with {\sc WiHaul} (observe flow $f_4$ in Fig.~\ref{fig:topo2thrput} with {\sc WiHaul} and BinDLMAC).

\begin{table}
\scriptsize
\begin{center}
 \begin{tabularx}{\columnwidth}{c|p{3.25em} p{4em} p{3.25em} p{4.5em} p{3.75em}} 
 \toprule 
\diaghead{\small Topology Sch}{Topology}{Scheme} &DLMAC &BinDLMAC &MDMAC &OptMDMAC &{\textsc {WiHaul}} 
\\ \hline 
1 & -28.912	&-25.658	&-12.186	&-14.75	&\textbf{-10.289}\\
2 & -31.783	&-33.605	&-11.542	&-13.848	&\textbf{-9.2215}\\
3 &-20.174	&-18.116	&-9.7012	&-10.826	&\textbf{-8.0635}\\
4 &-20.084	&-28.705	&-12.31	&-11.517	&\textbf{-9.3743}\\
\bottomrule
 \hline
\end{tabularx}
 \caption{$\mathcal{M}_\beta$ measure of max-min fairness as derived in (\ref{eq:fairnessmeasure}) following Lan's model \cite{lan:2010}. Simulation results.}
 \label{tab:lanfair}
 \end{center}
 \vspace*{-1em}
\end{table}

\vspace*{1em}
\hrule
\vspace*{0.25em}

\noindent\textbf{Finding: {\sc WiHaul} does not unnecessarily penalise flows that terminate/originate further away from gateways.} 

\vspace*{0.25em}
\hrule
\vspace*{1em}

Note in Figs.~\ref{fig:topo3thrput}--\ref{fig:topo4thrput} that with {\sc WiHaul} all flows achieve the same throughput for topologies \mbox{3--4}, unlike with DLMAC, MDMAC, and their variations, which largely favour flows terminating closer to the gateway and penalise those with end-points multiple hops away. (Opt)MDMAC is less prone to such behaviour, though has the disadvantage of requiring appropriate configuration of the slot size, which is impractical. Nonetheless, although the `optimised' MDMAC version performs relatively well overall, it still carries unfairness, as e.g. with this scheme flow $f_7$ in the third topology attains nearly half the throughput provided by {\sc WiHaul} (Fig.~\ref{fig:topo3thrput}).

To examine closer the fairness properties of all schemes, in Fig.~\ref{fig:gini} we plot the Gini coefficients corresponding to the flow rate allocations each of these yields in the 5 topologies considered. Recall the Gini coefficient gives a numerical representation of inequality, with a lower value corresponding to a fairer allocation. 
Observe that although these values depend on the network topology, number of flows, and link rates,  {\sc WiHaul} outperforms the existing schemes, being in particular considerably more fair than the DLMAC variants. Precisely, the Gini coefficients when the network operates with BinDLMAC range between 0.2 and 0.5 and are the highest in all 4 topologies. DLMAC performs marginally better, while (Opt)MDMAC yields Gini coefficients in the  0.1--0.3 range. Our proposal leads to the lowest Gini coefficients in all topologies (0.004--0.2), being substantially less unfair than the others.
These properties are further confirmed by the results we give in Table~\ref{tab:lanfair}, which shows the fairness measure as derived in (\ref{eq:fairnessmeasure}) for our approach and the benchmarks considered. Indeed $\mathcal{M}_{\beta}$ is up to 5$\times$ higher with our approach, which also indicates {\sc WiHaul} ensures superior performance for the smallest flow, yet remains fair to the others. 
 
We conclude that existing decentralised approaches bias against flows with longer hop-distance and/or inferior link rates; in contrast, the proposed {\sc WiHaul} not only achieves more fair partitioning of resources among all traffic flows, but also higher throughput for the smallest flow and overall higher mean throughput performance.
This has important practical implication on cellular backhauls where {\sc WiHaul} could provide superior and more homogeneous service guarantees to users.


\subsection{Dynamic Conditions}
\label{sec:dynam}
\begin{figure}[!t]
	\begin{center}
 	\includegraphics[width=1.0\columnwidth]{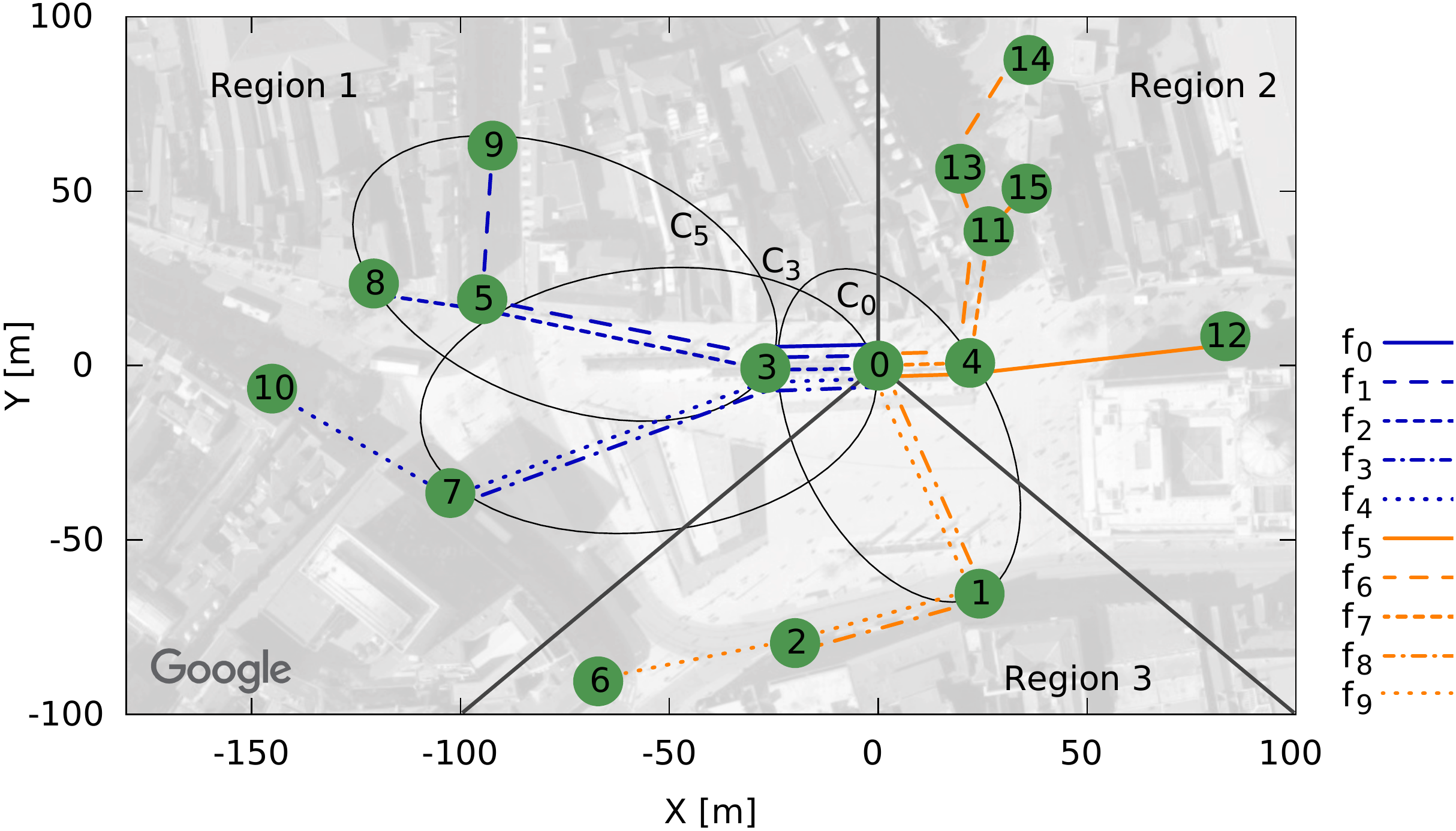}
 	\caption{Lamppost small cell backhaul deployment in Nottingham city centre, operating on mm-wave links. Geographic information extracted from open data set \cite{website:opendataNottin}; backhaul carries 10 aggregate flows; cliques of interest circled; topology `partitioned' into 3 regions.}
 	\label{fig:nottintopology}
	\end{center}
\end{figure}

Next we undertake an in-depth analysis of {\sc WiHaul}'s operation, investigating the impact of network dynamics in terms of link quality, flow demand, and routing strategy, on the airtime allocation and end-to-end performance. For this we envision a lamppost based deployment in the Old Market Square of Nottingham as shown in Fig.~\ref{fig:nottintopology}, which we obtain from a publicly available data set~\cite{website:opendataNottin}.  
This topology consists of 16 base stations (STAs) that communicate over mm-wave links and we envision 10 aggregate flows from the gateway (STA0). Also shown in the figure are three cliques of interest and, for ease of explanation, we consider the deployment as `partitioned' into three regions. 

\vspace*{0.45em}
\hrule
\vspace*{0.2em}

\noindent\textbf{Finding: Max-min fair backhauling requires a non-trivial partitioning of the available airtime resources, which depends on the demand of each flow, the paths traversed, and the capacities of the links these comprise.}

\vspace*{0.25em}
\hrule

\begin{figure*}[!th]
	\begin{center}
		\begin{subfigure}{0.49\textwidth}
			\includegraphics[width=\textwidth]{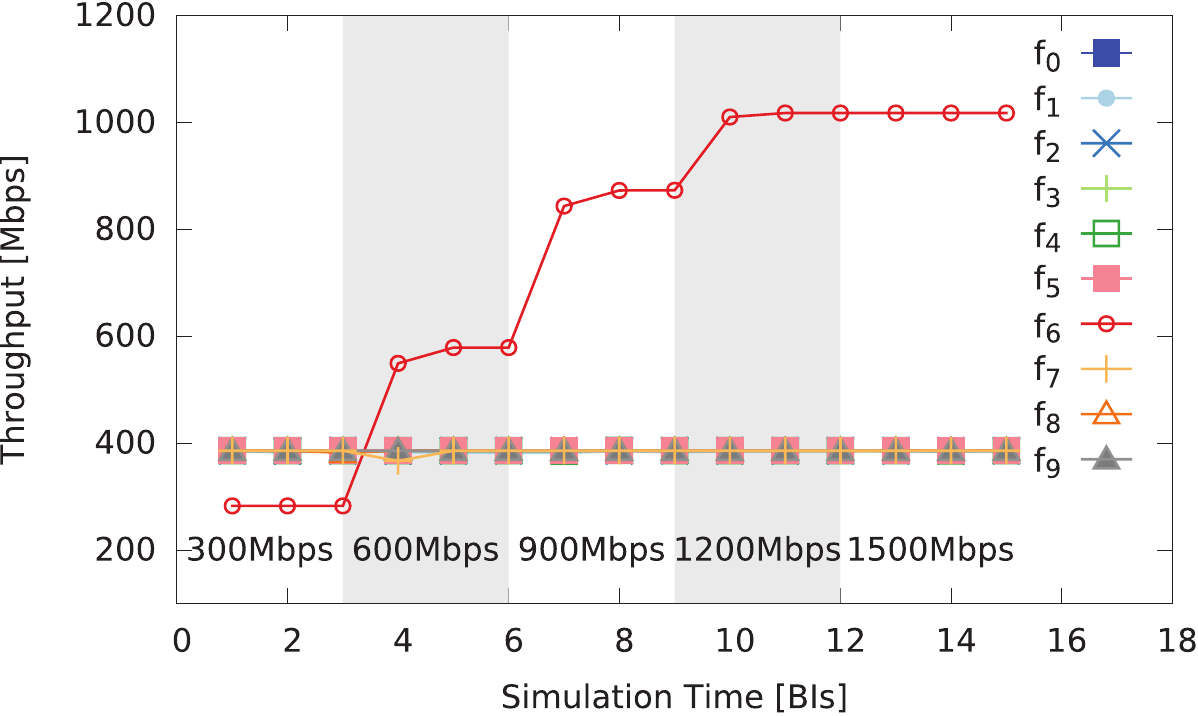}
			\caption{Time evolution of average flow throughputs.}
			\label{fig:dmdthroughput}
		\end{subfigure}
		\begin{subfigure}{0.49\textwidth}
			\includegraphics[width=\textwidth]{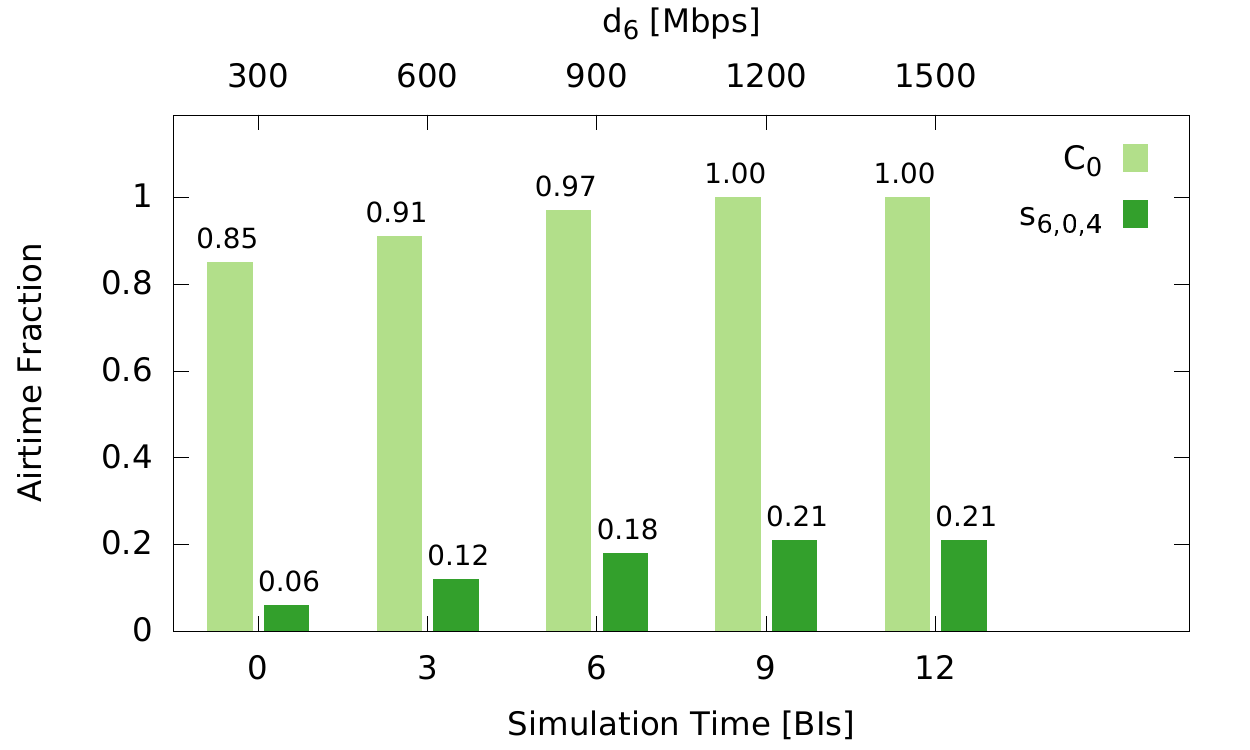}
			\caption{Airtime allocation for $f_6$, i.e $t_{6,0,4}$, and total airtime in $C_0$.}	
			\label{fig:dmdtime}
		\end{subfigure}
		\caption{Throughput performance and distribution of resources with {\sc WiHaul} in the topology in Fig.~\ref{fig:nottintopology} as the demand of $f_6$ increases from 300Mbps to 1.5Gbps, in 300Mbps steps every 3~BIs (in plot/top labels), while the demand of others remains at 400Mbps. Simulation~results.}
		\label{fig:dmdincreasing}
	\end{center}
\end{figure*}

\subsubsection{Demand Variation}
We first examine a scenario where the demand of a single flow (i.e. $f_6$ originating at STA0 and terminating at STA14) grows from 300Mbps to 1.5Gbps, while that of the others remains fixed to 400Mbps. Our goal is to understand how this impacts on airtime allocations and verify that the rates of the smallest flows are unaffected. We illustrate the results of this experiment in Fig.~\ref{fig:dmdincreasing}, where we plot (a) the time evolution of the individual throughputs and (b) the fraction of airtime allocated to $f_6$ on link $l_{0,4}$, as well as the total airtime allocated in Clique $C_0$, which constrains $f_6$.

Observe that the throughput of $f_6$ increases with demand, up to 1Gbps, when the clique constraint is activated (total airtime in $C_0$ reaches~1)
and the throughput is capped despite further growth in demand. As intended, the throughput of the remaining flows stays at 400Mbps, which indicates their demand is satisfied throughout. Note that the scheduling process is repeated every BI, link rate and demand updates are collected during BHIs, and it takes one BI duration for the demand increase to propagate through the network. 

To better understand the reasons behind these flow throughputs, we examine in Fig.~\ref{fig:dmdtime} the airtime utilisation in the bottleneck clique $C_0$ and the time allocated to the demand-varying flow, $t_{6,0,4}$. Observe that initially there exist sufficient resources to accommodate the entire demand of flow $f_6$; this holds for a demand up to 900Mbps, when $t_{6,0,4}$ is tripled. Further increasing this demand does not result in a throughput increase above 1Gbps. This is because our solution protects the remaining flows, which complies with the max-min fair allocation paradigm proposed. 


\begin{figure*}[!t]
	\begin{center}
		\begin{subfigure}{0.49\textwidth}
			\includegraphics[width=\textwidth]{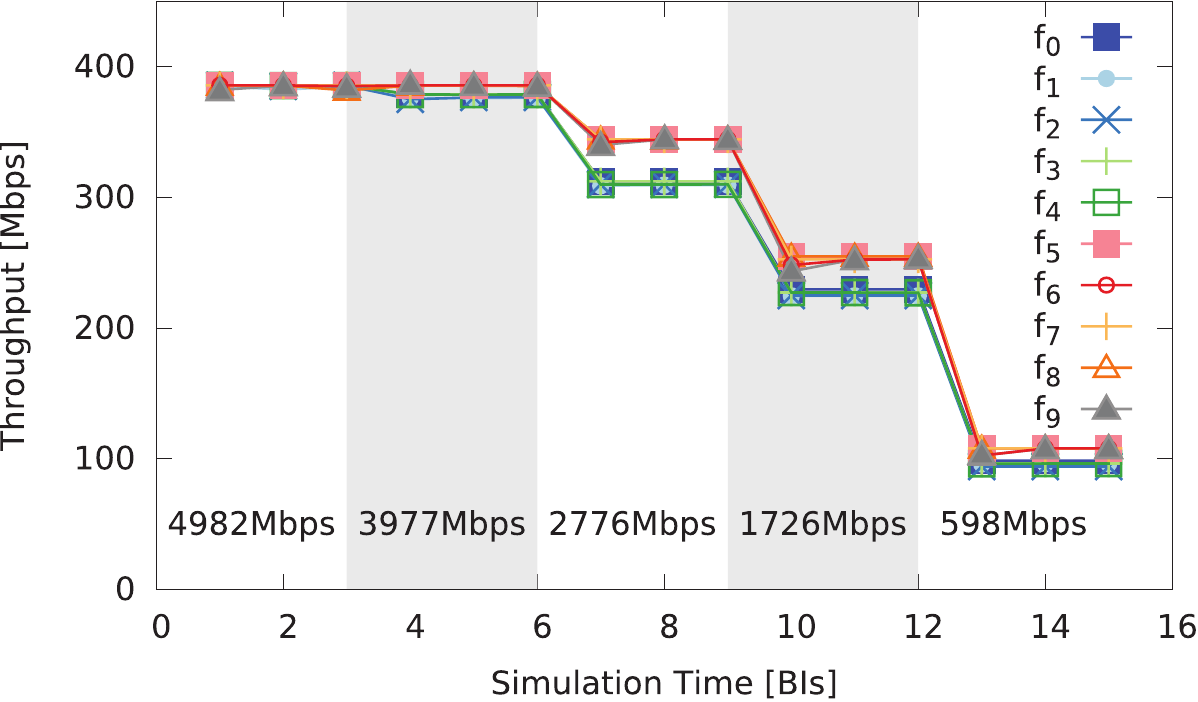}
			\caption{Evolution of flow throughputs.}
			\label{fig:sharedlinkthroughput}
		\end{subfigure}
		\begin{subfigure}{0.49\textwidth}
			\includegraphics[width=\columnwidth]{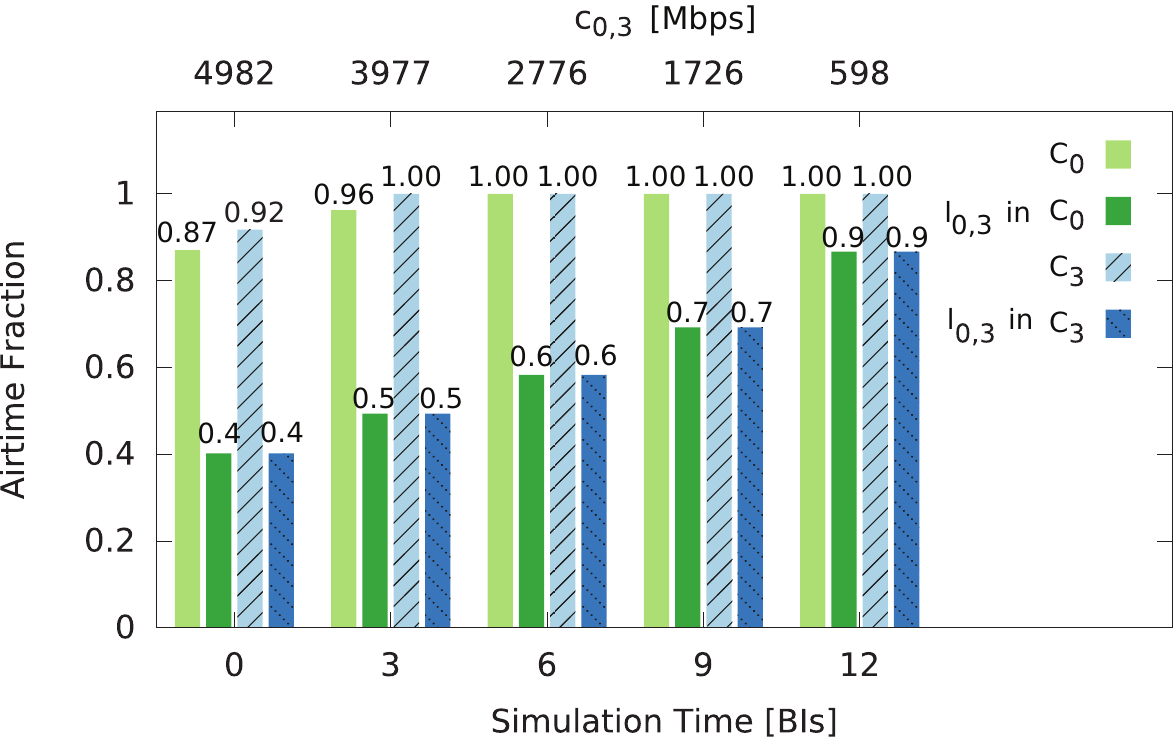}
			\caption{Airtime on link $l_{0,3}$ in cliques $C_0$ and $C_3$; total airtime in these.}
			\label{fig:sharedlinktime}
		\end{subfigure}	
		\caption{Flow throughputs and airtime fractions allocated to flow segments traversing $l_{0,3}$ in both cliques, as $l_{0,3}$ degrades. Flow demands remain at 400Mbps. MCS used with different link conditions labelled on white/shaded areas or shown as top x-ticks. Simulation results.}
		\label{fig:sharedlink}
	\end{center}
	\vspace*{-1em}
\end{figure*}

\subsubsection{Shared Link Degradation} 
Next we examine the impact of link quality variation on the performance of all flows traversing such a link, when max-min fair allocation is performed. To this end, we simulate different degrees of link blockage between STA3 and STA0 (i.e. $l_{0,3}$), which results in signal attenuation between 5dB and 20dB. As a result, the MCS employed is reduced from 4.9Gbps to 598Mbps, to preserve link reliability. In this scenario, we assume the bit rates of the other links remain constant and the demand of all flows is 400Mbps.

Fig.~\ref{fig:sharedlink} illustrates the results of this experiment, where we measure (a) the individual flow throughputs and (b)~the total time utilisation in cliques $C_0$ and $C_3$, as well as the sum of airtime fractions allocated to all flow segments traversing $l_{0,3}$, from the perspective of these cliques. Note that the airtime allocation on $l_{0,3}$ is effectively fixed under each link quality condition, but it may well represent different fractions from the cliques' perspectives. When the link quality is high (i.e. $c_{0,3}=4.982$Gbps), the total airtime consumption in $C_0$ and $C_3$ is below 1, hence all flows are satisfied. This is indeed confirmed by the flow throughputs shown in Fig.~\ref{fig:sharedlinkthroughput}. Subsequently, when a 5dB attenuation is introduced at the third BI, the throughputs of flows $f_0$--$f_4$ drop slightly, while those of $f_5$--$f_9$ remain satisfied. That is because $C_0$ still has sufficient resources (airtime consumed sums to 0.96), while the $C_3$ clique constraint becomes active (airtime reaches~1). This can be observed indeed in Fig.~\ref{fig:sharedlinktime}, where we also see that the total airtime allocated on link $l_{0,3}$ increases from 0.4 to 0.5 in both cliques, as a result of signal degradation.

Further attenuation on link $l_{0,3}$ (yielding 2.776Gbps bit rate), leads to the activation of the $C_0$ constraint (observe in \ref{fig:sharedlinktime} that the total airtime in clique $C_0$ reaches 1), and consequently to a decrease in the throughput of all flows. However, as $C_3$ becomes constrained before $C_0$, flows $f_0$--$f_4$ attain slightly lower (approx. 30Mbps) throughput than $f_5$--$f_9$. Lastly, this performance gap shrinks as link $l_{0,3}$ degrades further (BI 9 onward) and additional degradation would completely close the gap to meet the max-min fairness criterion. Meanwhile, the total time consumed by link $l_{0,3}$ to transport all flows is increasing to as much as 0.9 at the end of the simulation (see BI 12 in Fig.~ \ref{fig:sharedlinktime}).

We conclude that degradation of an intensively shared link (and clique) has a significant impact on the throughput performance of the entire network. Nevertheless, {\sc WiHaul} guarantees max-min fair allocation of the flow rates.
 
\subsubsection{Heterogeneous Demands and Cascaded Cliques} 
\begin{figure*}[!t]
	\begin{center}
		\begin{subfigure}{0.495\textwidth}
			\includegraphics[width=\textwidth]{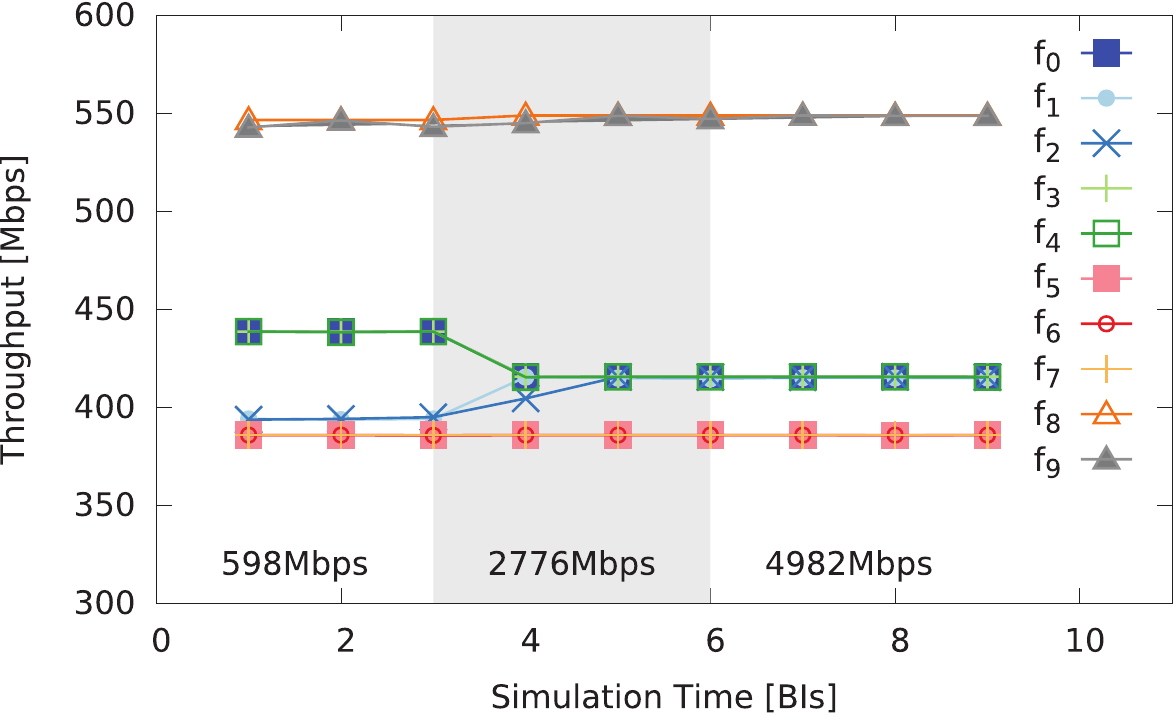}
			\caption{Flow throughput evolution.}
			\label{fig:terminallinkthroughput}
		\end{subfigure}
		\begin{subfigure}{0.495\textwidth}
			\includegraphics[width=\columnwidth]{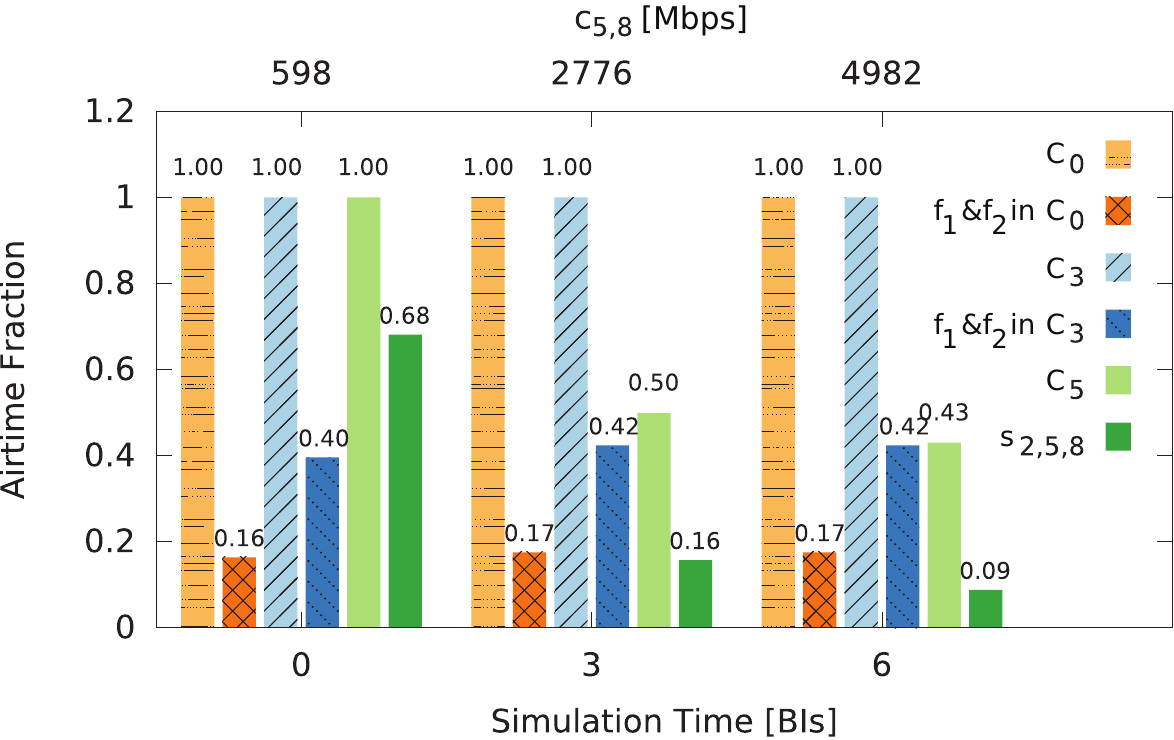}
			\caption{Total airtime allocated in cliques $C_0, C_3$ and $C_5$; time assigned to $f_1$ and $f_2$ in $C_0$ and $C_3$; time allocated to $f_2$ on $l_{5,8}$, i.e. $t_{2,5,8}$.}
			\label{fig:terminallinktime}
		\end{subfigure}
		\caption{Throughput performance and resource partitioning as $c_{5,8}$ increases from 598 to 4,982Mbps. Flow demands in region 1, 2 and~3 are 500, 400, and 600Mbps respectively. Shaded and label areas/x-ticks at the top correspond to bit rates on $l_{5,8}$ as link quality changes. Simulation results.}
	\end{center}
	\label{fig:terminallink}
\end{figure*}

In what follows we consider more complex circumstances where the demands of flows in regions 1--3 as shown in Fig.~\ref{fig:nottintopology} are 500, 400, and 600Mbps respectively, while the quality of link $l_{5,8}$ varies. Signal attenuation decreases and capacity grows from 598 to 4,982Mbps on this link after every third BI. As $l_{5,8}$ only carries flow $f_2$, we investigate in Fig.~\ref{fig:terminallinktime} the changes in time allocation within all the cliques that $f_2$ traverses, i.e. $C_5$, $C_3$, and $C_0$, and show the time evolution of individual flow throughputs in Fig.~\ref{fig:terminallinkthroughput}. 

Note that as $c_{5,8}$ increases, more airtime is made available for both $f_1$ and $f_2$, as they share the same clique $C_5$. In effect, the constraint of this clique is removed (total airtime consumption drops from 1 to 0.5) and this also impacts on the flows with which $f_1$ and $f_2$ share cliques $C_3$ and $C_0$, i.e. $f_0$, $f_3$ and $f_4$. Precisely, the throughput of these drops to 415Mbps after the third BI. As the quality of $l_{5,8}$ further increases, the total airtime allocated to $f_2$ on this link, i.e. segment $s_{2,5,8}$, decreases, though the 
flows in region 1 are together constraint by $C_3$. This confirms the proposed
max-min fair allocation strategy ensures $f_2$ is not allocated more resources in cliques $C_3$ and $C_0$, as this would come at the cost of a decrease in the throughput of flows $f_0$, $f_3$, and $f_4$. Lastly, observe that the throughput of the other flows remains unaffected, as the demand of $f_5, f_6, \text{and} f_7$
 is the smallest among all (i.e. 400Mbps) and changes in $c_{5,8}$ do not affect clique $C_0$, which is shared by all flows.

\begin{figure*}[!t]
	\begin{center}
	    \begin{subfigure}{\textwidth}
	    \centering
            \includegraphics[trim=0 0 80 0, clip, width=0.9\textwidth]{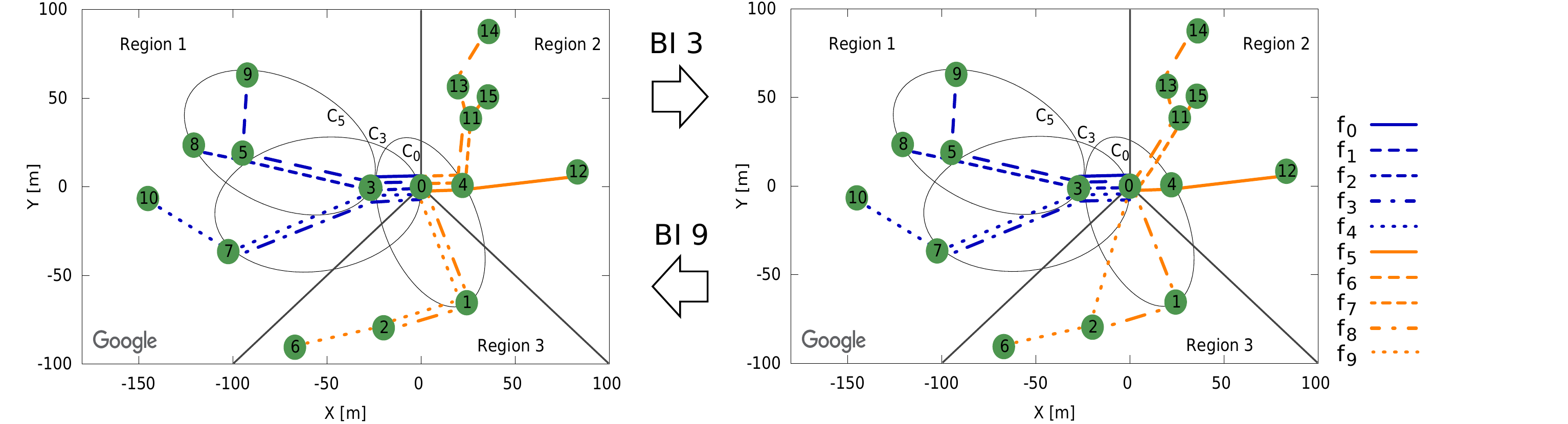}
            \caption{Changes in routing topology taking place at BI 3 and BI 9. At BI 3, flows $f_6$, $f_7$, and $f_9$ are rerouted, while the paths of $f_5$ and $f_9$ remain unchanged. At BI 9, the routing topology reverts back to the initial one.}
            \label{fig:s7transition1}
        \end{subfigure}
		\begin{subfigure}{0.495\textwidth}
			\includegraphics[width=\textwidth]{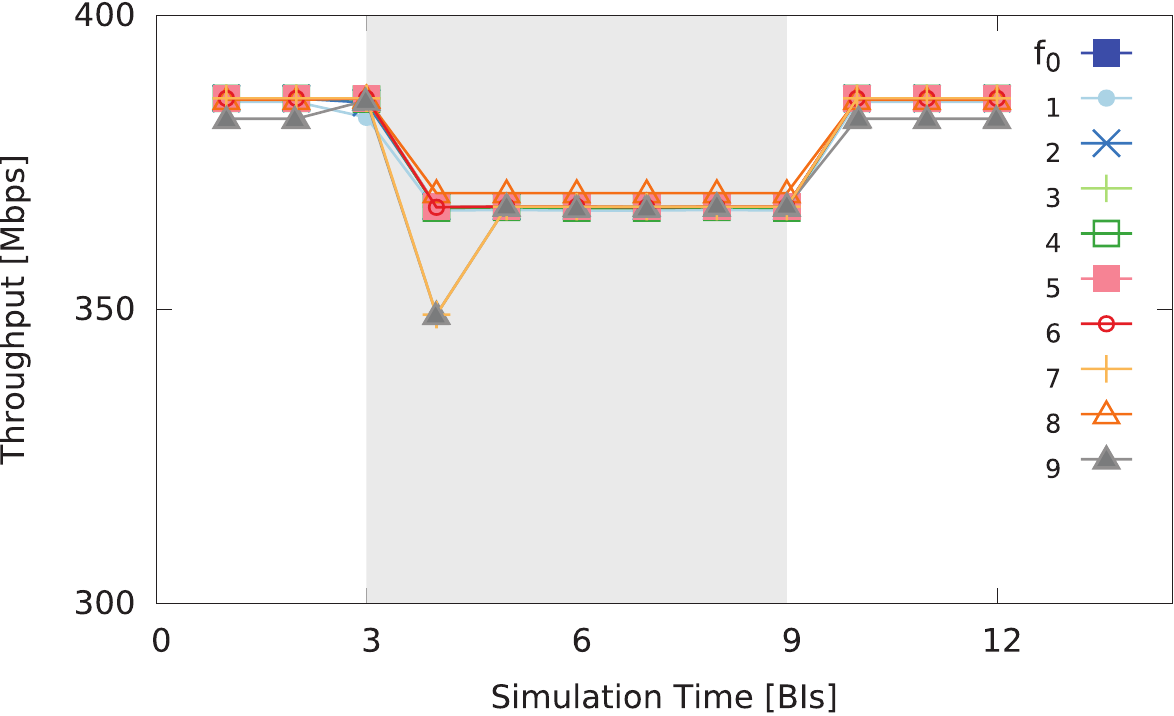}
			\caption{Time evolution of flow throughputs.}
			\label{fig:routingthroughput}
		\end{subfigure}
		\begin{subfigure}{0.495\textwidth}
			\includegraphics[width=\columnwidth]{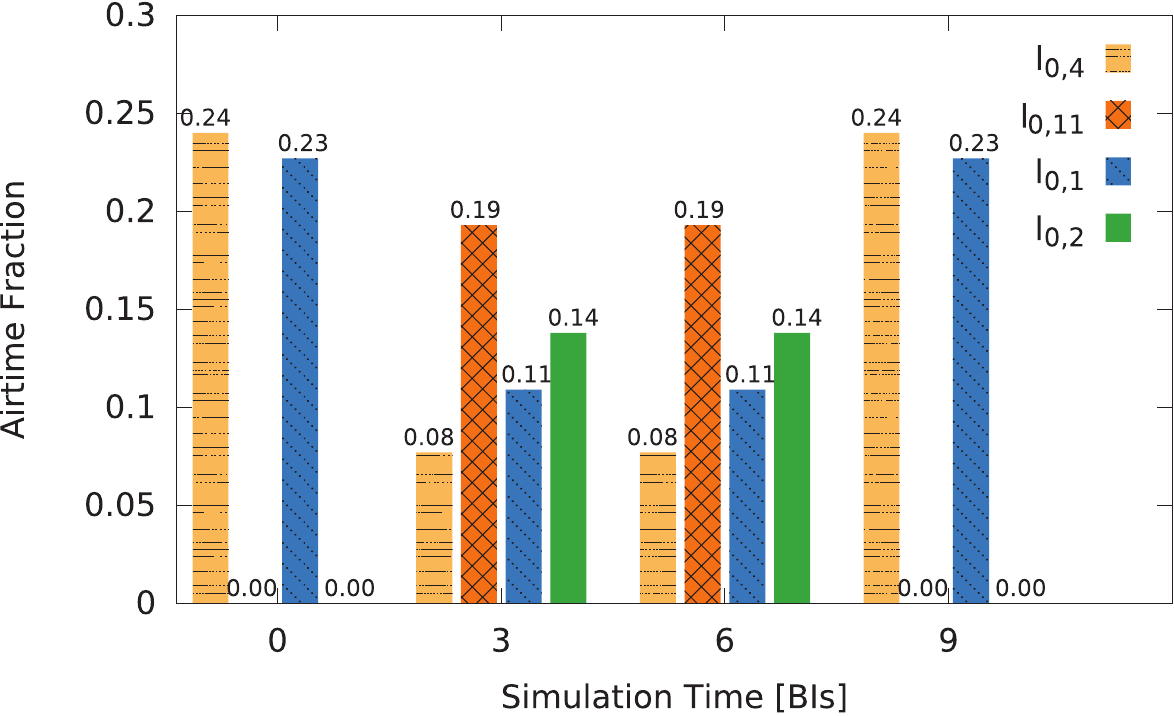}
			\caption{Airtime allocated on links $l_{0,4}$, $l_{0,11}$, $l_{0,1}$, and $l_{0,2}$.}
			\label{fig:routingtime}
		\end{subfigure}
				
		\caption{Throughput performance and resource partitioning as the routing topology changes. Flow demands for all are 400Mbps respectively. Shaded area corresponds to routing changes. Simulation results.}
	\end{center}
	\label{fig:routing}
	\vspace*{-1em}
\end{figure*}

\subsubsection{Dynamic Routing}
Next we investigate the impact of route changes on the airtime allocation and end-to-end throughput performance, when the backhaul is managed with the proposed {\sc WiHaul} solution. To this end, we consider a situation where part of the traffic traversing links $l_{0,4}$ and $l_{0,1}$ in the topology depicted in Fig.~\ref{fig:nottintopology}, i.e. flows $f_6$, $f_7$, and $f_9$ are rerouted to STA2 and STA11 (i.e. no longer traverse STA1 and STA4), while the routes followed by $f_5$ and $f_8$ remain unchanged. After 6 beacon intervals, the initial routing topology is restored. We illustrate these changes in Fig.~\ref{fig:s7transition1}.
Such routing changes can happen due to link blockage, buffer overflows, or other routing decisions made by a routing algorithm running at the networking layer.

We show in Fig.\ref{fig:routingthroughput} the end-to-end throughput dynamics for all flows, as a result of these route changes, and in Fig.~\ref{fig:routingtime} the corresponding time allocation on links $l_{0,4}$, $l_{0,11}$, $l_{0,1}$, and $l_{0,2}$. Observed that {\sc WiHaul} reacts fast by re-allocating the airtime resources and the network throughput is only marginally affected. Flows $f_9$ and $f_7$ experience a 35Mbps drop at BI 4 due to the fact that packets buffered at STA4 and STA1 are partially dropped when the routes change, but the throughput recovers in the following BI. These results also confirm that {\sc WiHaul} will not unnecessarily penalise flows traversing more hops. In particular, when the routes change and the number of hops traversed by flows $f_6$, $f_7$, and $f_9$ decreases, after re-computing rates with the max-min criterion, their throughput is actually reduced, due to the fact that the clique $c_0$ consists of link segments, $l_{0,11}$ and $l_{0,2}$ that observe lower capacity as compared to links on the original paths, i.e. $l_{0,4}$ and $l_{0,1}$.

As expected, the time allocated on links $l_{0,4}$ and $l_{0,1}$ is reduced by approximately $2/3$ and $1/2$ when the routes of $f_6$, $f_7$ and $f_9$ change. Meanwhile, the time fractions allocated to $l_{0,11}$ and $l_{0,2}$ increase from 0 to 0.19 and 0.14, which are both more than the amount reduced in $l_{0,4}$ and $l_{0,1}$. This is because the capacity of the new links employed by the new routes are lower than those on the initial routes.

\subsection{Secondary Interference}
\label{sec:2ndintf}
In this subsection we examine the potential impact of secondary interference, showing how {\sc WiHaul} can overcome this by constructing cliques that capture such circumstances and avoiding their simultaneous activation during scheduling. We also discuss the complexity cost incurred when accounting for such secondary interference.

\begin{figure}[t!]
	\begin{center}
 	\includegraphics[width=1.0\columnwidth]{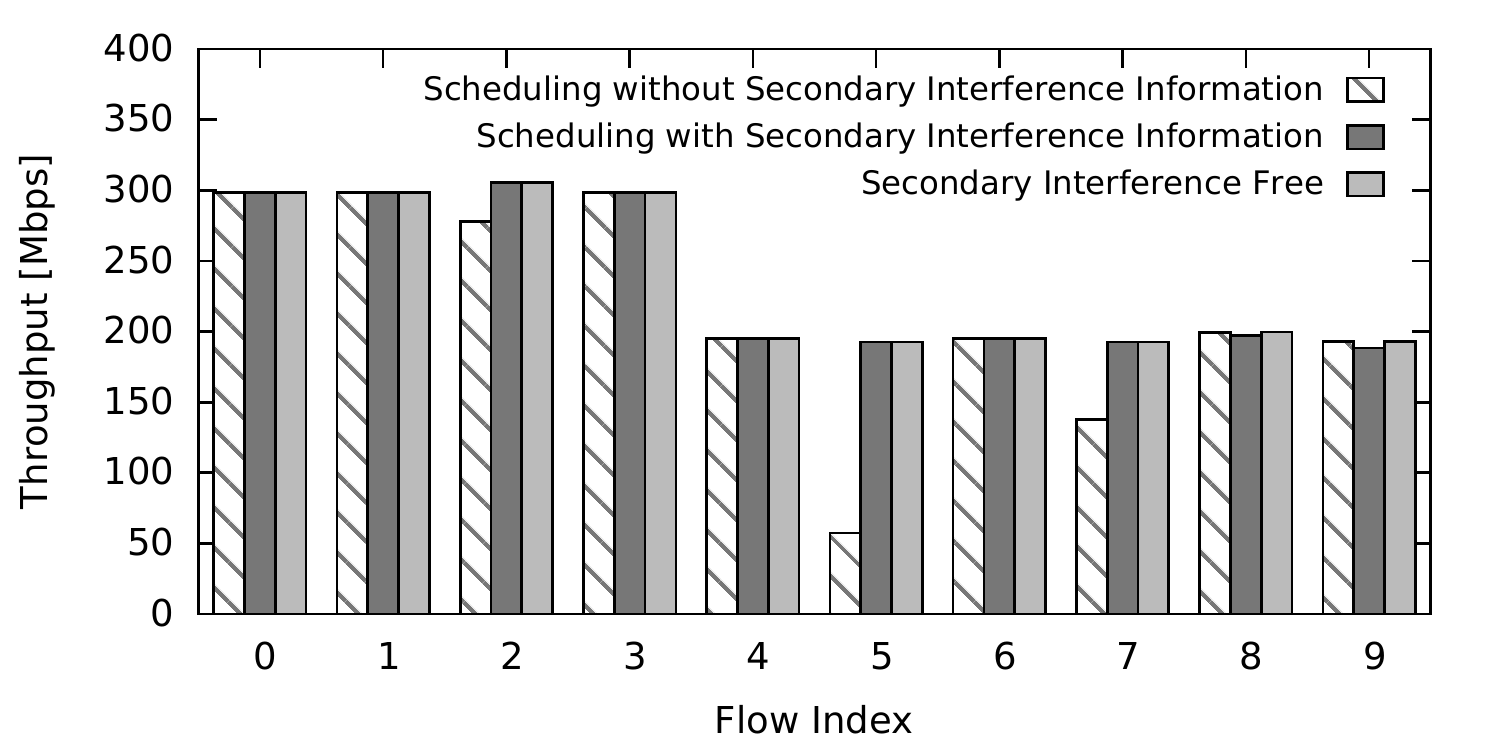}
 	\caption{The existence of secondary interference can degrade the throughput of flows traversing interfering links (i.e. Flows 2, 5, and 7). Taking into account the potential secondary interference, {\sc WiHaul}'s mitigates this effect.}
 	\label{fig:interf}
	\end{center}
\end{figure}

We simulate again the topology shown in Fig.~\ref{fig:topo2map}, where some links may interfere with each other when their TX/RX beam pairs are aligned. Specifically, when STA7 is receiving from STA0 and STA5 is transmitting to STA14, STA7 experiences secondary interference as the power of the signal it receives from STA5 has a level of $-62.7$dBm. STA5 can suffer the same if receiving from STA14 and STA7's transmission to STA0 happens at the same time. Moreover, STA13's transmission to STA10 will interfere with STA14, if this is beam-switched to STA12 for reception, and vice versa.
Fig.~\ref{fig:interf} illustrates the individual flow throughput averaged over 20 BIs when 1) secondary interference exists but the scheduling ignores this; 2) secondary interference exists and {\sc WiHaul} incorporates this information when performing scheduling; and 3) the system is free of secondary interference.
Observe that flows traversing the interfering links, i.e. $f_2$, $f_5$, and $f_7$, experience 50Mbps, 150Mbps, and respectively 70Mbps throughput degradation when schedules are assigned without accounting for such interference. When {\sc WiHaul} employs this knowledge for transmission coordination, the cliques are constructed such that none of the potentially interfering links are active simultaneously, regardless of whether this is due to secondary interference. As a results, the flow throughputs obtained when secondary interference is accounted for are virtually the same as those achieved in the idealistic case of the topology being free of secondary interference (given perfect beam shapes and pseudo-wired communication).

Given that secondary interference is in most cases marginal, and only 4 out of the total of 47 links in the simulated topologies shown in Fig.~\ref{fig:4topos} experience secondary interference, it is worth understanding the computational cost of scheduling with secondary interference in mind. Each group of interfering links forms a clique and a link with lower priority in the scheduling hierarchy (see Sec.~\ref{sec:wihaulprotocol}) will have to store the time slots used by the links with higher priority in the hierarchy, which introduces $n_{sche}$ iterations. Hence, accounting for secondary interference will increase the computation complexity of {\sc WiHaul} proportionally with the number of links that may interfere with each other if active simultaneously.

\subsection{Real-Time Traffic}
\begin{figure}[t]
\begin{center}
 \includegraphics[width=0.8\columnwidth]{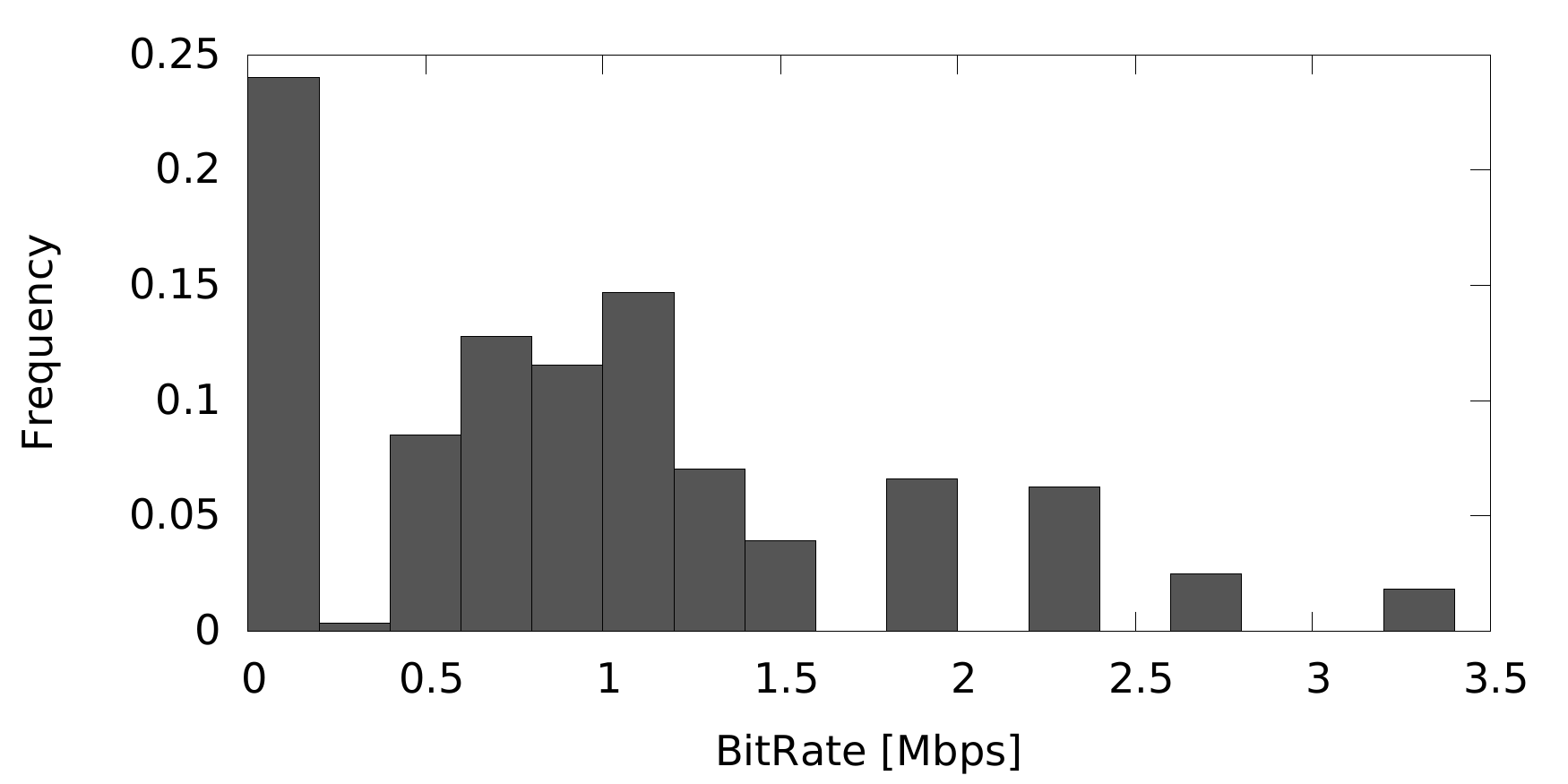}
 \caption{Distribution of DASH flow bit rates measured in New York City as reported in \cite{nyupoly-video-20140509} and used here for the evaluation of {\sc Wihaul}.}
 \label{fig:dashbitrate}
  \end{center}
\end{figure}

We continue our evaluation of {\sc WiHaul} by conducting experiments with real-time traffic potentially subject to latency constraints. We are particularly interested in the delay packets experience while traversing multi-hop mm-wave backhauls, where cascaded queues could have a negative impact on user experience. To this end, we emulate dynamic adaptive streaming over HTTP (DASH) by extracting meta-data from mobile traffic traces collected in New York City~\cite{nyupoly-video-20140509}. We replay 100 such video sessions in parallel towards different base stations (download) in the topology shown in Fig.~\ref{fig:nottintopology}. The distribution of the session bit rates is shown in Fig.~\ref{fig:dashbitrate}, where observe that individual bit rates vary between 100Kbps and 3.4Mbps.

\begin{figure*}
	\begin{center}
		\begin{subfigure}{0.49\textwidth}
			\includegraphics[width=\textwidth]{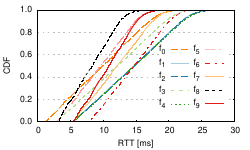}
			\caption{Round trip time distribution.}
			\label{fig:dashrtt}
		\end{subfigure}
		\begin{subfigure}{0.49\textwidth}		
			\includegraphics[width=\textwidth]{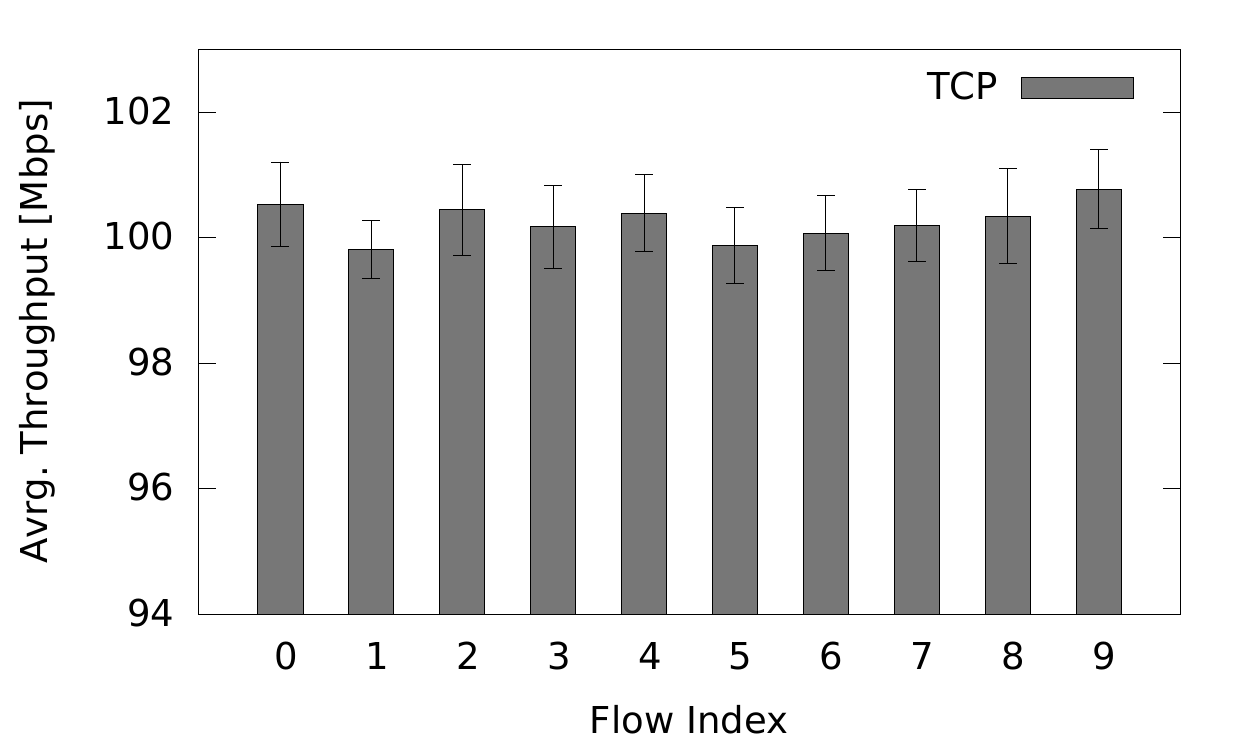}
			\caption{Average aggregate flow throughput.}
			\label{fig:dashthrpt}
		\end{subfigure}
	\caption{CDF of packet RTTs for the aggregate flows (each consisting of 100 HTTP sessions following the bit rate distribution in Fig.~\ref{fig:dashbitrate}) in the Nottingham topology shown in Fig.~\ref{fig:nottintopology}, and their average throughputs. Simulation results.}
	\end{center}
	\vspace*{-1em}
\end{figure*}

Under these circumstances, we measure the packet round-trip-time (RTT) for each aggregate flow over 30 seconds, as well as the average throughputs. 
We plot the RTT experienced by TCP segments in Fig.~\ref{fig:dashrtt}, where observe this is below 30ms, with median values for all aggregates falling between 8 and 15ms. This complies with the NGMN Alliance specifications for end-to-end delay (20ms) in small cell backhauls \cite{ngmn:2012}.
As expected, RTTs are proportional to the number of hops traversed, however, their distribution also depends on how frequently they are served. Precisely, note that the slope of the CDFs decreases with the number of aggregates traversing the first hop from the gateway and thus the latency in different regions is only scaled up by the number of hops each aggregate traverses. For instance, flows $f_1, f_2,$ and $f_4$ are 3 hops away from the gateway (STA0) and share $l_{0,3}$ with $f_0$. As such, the RTTs they experience are identical (overlapping curves). Flow $f_9$ also traverse 3 hops, but only shares $l_{0,1}$ with $f_8$, hence their RTT distributions start at $\sim$5ms, but quickly diverge (medians 11 and respectively 15ms).

Turning attention to aggregate flow throughputs, we show the average and 95\% confidence intervals of this metric in Fig.~\ref{fig:dashthrpt}. We see that overall performance is homogeneous (despite flows traversing different number of hops and experiencing different link rates), fluctuating around 100Mbps for each aggregate. Note that in this scenario all flows are satisfied and cliques are not constrained.

\subsection{Runtime Performance}
\label{sec:runtime}

Lastly, we examine the runtime convergence of {\sc WiHaul}'s progressive filling routine, to understand the practical feasibility of executing this algorithm periodically in order to perform airtime allocation. To this end, we take again the Nottingham topology depicted in Fig.~\ref{fig:nottintopology}, as this has a reasonably large number of nodes (i.e. 14) and aggregate flows traversing it (i.e. 9), which directly impact on the complexity. We measure the total time required by an off-the-shelf workstation, equipped with an Intel Core i5-4570 CPU clocked at 3.20GHz, to complete the execution of the progressive filling. For these measurements, we consider all flows have equal demands that range between 100Mbps and 2Gbps, with 50Mbps increments. For each case, we set the step length of the progressive filling algorithm to 10Mbps, execute this algorithm 100 times, and compute the mean runtime with 95\% confidence intervals.
\begin{figure}
    \centering
    \includegraphics[width=\columnwidth]{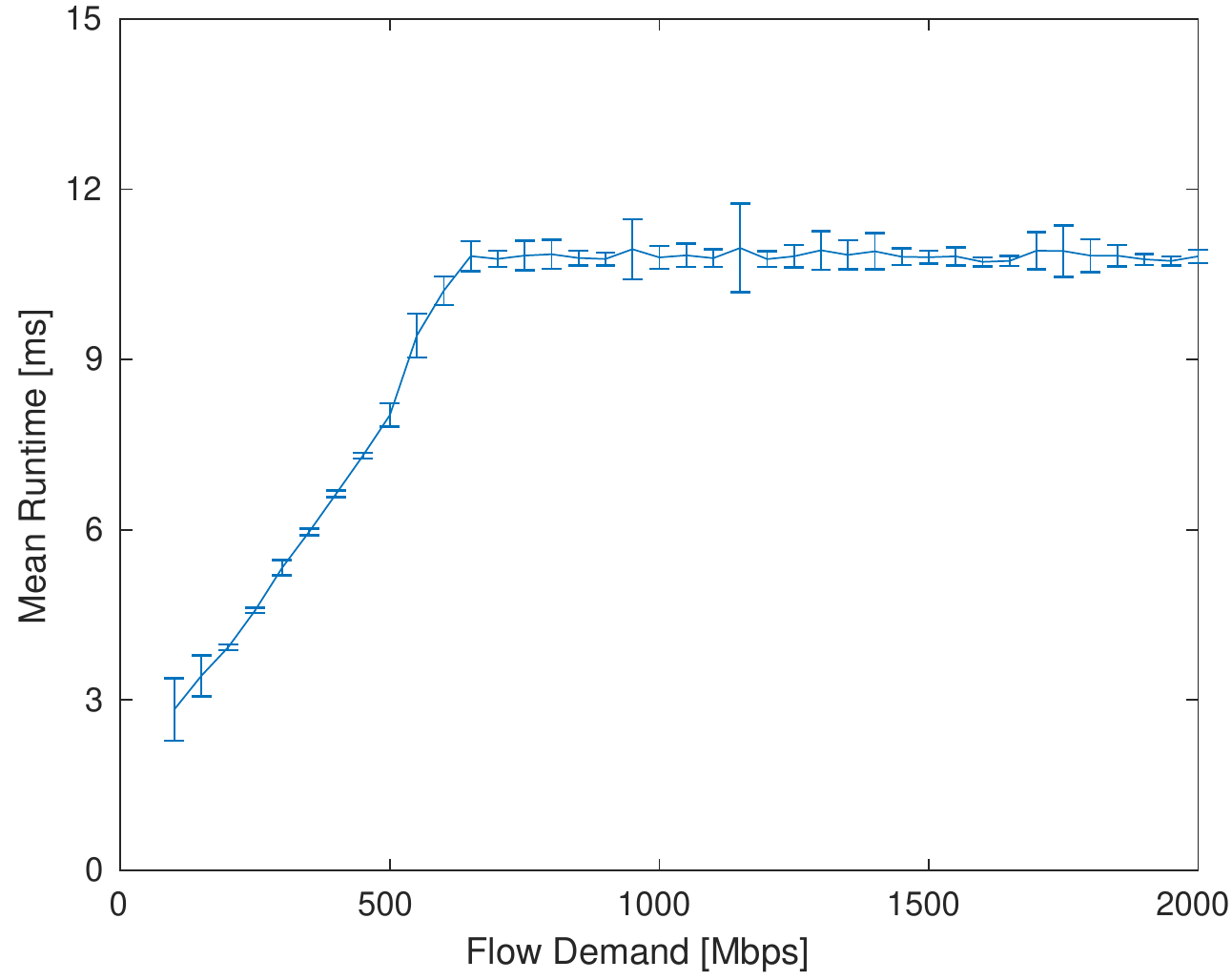}
    \caption{Progressive filling runtime performance as flow demands increase, in the topology shown in Fig.~\ref{fig:nottintopology}. Simulation results.}
    \label{fig:my_label}
    \vspace*{-1em}
\end{figure}

Observe that our solution converges within a number of steps that, as long as clique constraints are not met (which will eventually happen, given limited channel capacities), strictly depends on the demand and the step size. For this topology, the runtime will not increase beyond approximately 11ms as the demand exceeds 650Mbps. We argue that this cost is negligible if the algorithm is run e.g. every second, while the granularity can be increased if the airtime allocation routine is fed with the output of a traffic forecasting mechanism~\cite{Zhang:2018} and executed in anticipation of the expected traffic.

\section{Conclusions}
\label{sec:conclusions}
By supporting multi-Gbps link rates, mm-wave technology is becoming a promising enabler of wireless backhauling solutions in ultra-dense cellular deployments. Highly directional beamforming is mandatory to combat severe signal attenuation specific to these frequencies, though this gives rise to cumbersome terminal deafness issues that must be tackled to fully exploit the vast bandwidth  available. In this paper, we proposed {\sc WiHaul}, a network-wide airtime resource allocation and scheduling mechanism that works with TDM-based medium access protocols (including 3GPP 5G NR and IEEE 802.11ad), which explicitly guarantees inter-flow max-min fairness in mm-wave backhauls. We validated our solution over a broad range of dynamic network conditions and demonstrated via extensive simulations that {\sc WiHaul} achieves up to 5$\times$ higher measurable fairness as compared to existing mm-wave MAC proposals, improving up to five-fold the individual throughput of otherwise limited flows, while attaining superior overall network performance. Further, we demonstrated that the progressive filling routine we devise for airtime allocation completes within milliseconds and its complexity strictly depends on the highest flow demand. Lastly, our approach is able to meet the typical delay constrains of real-time applications.

\section*{Acknowledgements}
We thank Pablo Serrano for his valuable feedback that helped improving this manuscript. We also thank Nicol\`o Facchi for helping us extend the NS-3 simulator, which we used for evaluation purposes.

\begin{IEEEbiography}[{\includegraphics[trim=80 200 60 70,width=1in,height=1.25in,clip,keepaspectratio]{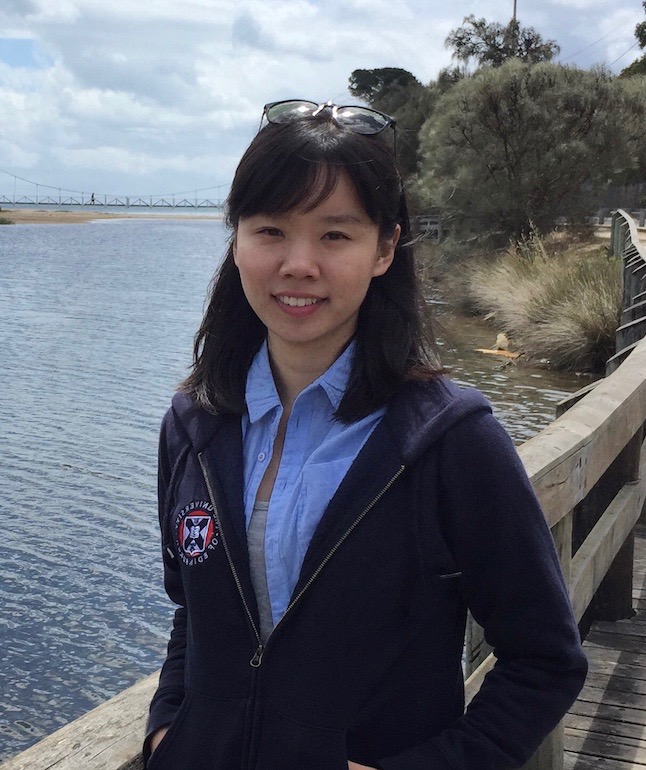}}] {Rui Li} received a B.Eng. degree from Northwestern Polytechnical University (2013) and an M.Sc. from University of Leicester (2014). She is currently a final year Ph.D. candidate in the School of Informatics at the University of Edinburgh, where she investigates resource allocation solutions for next generation mobile networks and machine learning applications to networking problems.
\end{IEEEbiography}

\begin{IEEEbiography}[{\includegraphics[width=1in,height=1.25in,clip,keepaspectratio]{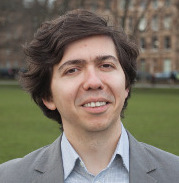}}] {Paul Patras} [SM'18, M'11] received M.Sc. (2008) and Ph.D. (2011) degrees from Universidad Carlos III de Madrid (UC3M). He is a Reader (Associate Professor) and Chancellor's Fellow in the School of Informatics at the University of Edinburgh, where he leads the Internet of Things Research Programme.  His research interests include performance optimisation in wireless and mobile networks, mobile intelligence, security and privacy.
\end{IEEEbiography}
\vfill
\end{document}